\documentclass[11pt]{article}
\usepackage[top=1.1in, left=1in, right=1in, bottom=1.1in]{geometry}
\usepackage{amsmath,amssymb,amsthm,amsfonts,verbatim}
\usepackage{enumerate}
\usepackage{url}
\usepackage{amsmath,amssymb}
\usepackage{graphicx}
\usepackage{color}
\usepackage{setspace}
\usepackage{xcolor}

\title{Modeling aggregation processes of Lennard-Jones particles via stochastic networks}

\author{Yakir Forman$^1$ and Maria Cameron$^2$}
\begin{document}

\maketitle
\abstract{
We model an isothermal aggregation process of particles/atoms interacting according to the Lennard-Jones pair potential 
by mapping the energy landscapes of each cluster size $N$ onto stochastic networks, computing transition probabilities {from} 
the network for an $N$-particle cluster to the one for $N+1$, and connecting these networks 
into a single joint network. The attachment rate is a control parameter.
The resulting network representing the aggregation of  up to 14 particles contains {6427} 
vertices.
It is not only time-irreversible but also reducible. To analyze its transient dynamics,
we introduce the sequence of the expected initial and pre-attachment distributions and compute them 
for a wide range of attachment rates and three values of temperature.
As a result, we find the {configurations most likely to be observed} 
in the process of aggregation 
for each cluster size. 
We examine the attachment process and conduct a structural analysis of the sets of local energy minima for 
every cluster size. We show that both processes taking place in the network, attachment and relaxation, 
lead to the dominance of icosahedral packing in small (up to 14 atom) clusters.
}

\section{Introduction}
\label{intro}
This work is inspired by the gap between theoretical studies of clusters of Lennard-Jones atoms and 
experimental works {in which} 
rare gas clusters are examined by means of electron \cite{exp1,exp2,exp3,exp4,exp5} or X-ray \cite{kakar} diffraction. The former 
achieved significant progress in understanding thermodynamics (e.g. Refs \cite{wales-thermo,mf}) and transition processes 
(e.g. Refs. \cite{wales0,picciani,wales-book,wales-landscapes}) of/in clusters of \emph{fixed} numbers of particles, while 
rare gas atoms self-assemble into clusters in experimental settings, and \emph{nothing prevents them from acquiring new atoms}.
Mass spectra measured in experimental work  \cite{exp1,exp2,exp3,exp4,exp5} 
provide a strong evidence that icosahedral clusters tend to form with small numbers of atoms, while 
face-centered cubic  (FCC) packing becomes prevalent for large clusters. The switch from icosahedral to 
FCC packing occurs somewhere in the range of cluster size {between} 
1500  
{and} 
$10^4$ atoms, while  the presence of FCC packing was detected  
in clusters of $N\ge 200$ atoms  \cite{kakar}.
What is the mechanism of this switch? 
Van de Waal hypothesized that this switch happens not due to rearrangement of atoms within clusters but
because faulty FCC layers start to grow on icosahedral cores \cite{wall}. 
Kovalenko et al \cite{kovalenko} inferred the structure of large rare gas clusters from experimental measurements and showed that
it was consistent with Van de Waal's conjecture.
\footnotetext[1]{ {\tt yakir.forman@mail.yu.edu}, Yeshiva University, 500 W 185th St, New York, NY 10033}
\footnotetext[2]{ {\tt cameron@math.umd.edu}, Department of Mathematics, University of Maryland, College Park, MD 20742}

\subsection{Intriguing facts about the self-assembly of free Lennard-Jones atoms}
The potential energy of a cluster of $N$ particles interacting according to the Lennard-Jones pair potential
written in reduced units is given by
\begin{equation}
\label{LJ}
V(\mathbf{r}_1,\ldots,\mathbf{r}_N) = 4\sum_{i=2}^N\sum_{j=1}^{i-1}(r_{ij}^{-12} - r_{ij}^{-6}),\quad r_{ij} = |\mathbf{r}_i - \mathbf{r}_j|,\quad \mathbf{r}_{i} = (x_i,y_i,z_i).
\end{equation}
The global energy minima for cluster sizes $2\le N\le 110$ are mostly achieved on
configurations with icosahedral packing; however,
for some special numbers of atoms, $N=38$, 75, 76, 77, 102, 103, and 104, the energy-minimizing configurations are non-icosahedral \cite{wales110}.
Some of them are highly symmetric. For example, the global minimum for $N=38$, a truncated octahedron with 
FCC atomic packing,
has the point group $O_h$ of order 48, i.e., there are 48 orthogonal transformations mapping the cluster onto itself. The global minimum for $N=75$,
a Marks decahedron, has point group $D_{5h}$ of order 20. Remarkably, the mass spectra graphs in  \cite{exp1,exp4} do not have 
prominent peaks at $N=38$ and $N=75$. 
On the other hand, the mass spectra in \cite{exp1,exp2,exp3,exp4,exp5} consistently exhibit peaks 
corresponding to the  clusters of the so-called magic numbers of atoms $N$ admitting complete icosahedra.
These numbers are: $N = 13$, 55, 147, 309, 561, etc. {The point group order of an icosahedron is 120. Evidently,  atoms tend to
self-assemble into highly symmetric complete icosahedra in experimental settings, 
while {they} seem to miss highly symmetric low-energy configurations based on other kinds of packing, at least for small numbers of atoms.}

\subsection{{ Choosing a model and an approach}}
\label{sec:model}
Intrigued by these facts, we undertook an attempt to understand the 
self-assembly of free Lennard-Jones particles (atoms) into clusters on the quantitative level by means 
of combined analytical and computational methods. {Most} previous theoretical studies of Lennard-Jones clusters dealt with
those of fixed numbers of atoms, i.e., atoms were {allowed neither} to fly away nor {to} join the cluster.
These works can be divided into two groups, full phase-space-based (e.g. \cite{mf,picciani}) and  network-based.
The latter approach was pioneered by Wales and collaborators \cite{LJ7,wales110,wales38,wales-book,wales-landscapes}. 
Their powerful computational tools for mapping energy
landscapes onto networks are based on the basin-hopping method \cite{wales110} and discrete path sampling \cite{wales0}.
Numerous networks representing energy landscapes of proteins (e.g. \cite{beta3}) and
clusters of particles interacting according to various pair potentials (e.g. \cite{helix,wales-short}) 
are available or advertised in {Wales's group's} 
web page \cite{web}.

{Variable size clusters were considered in a few earlier works as well.
The formation of low-energy minima of metallic clusters Ag$_{38}$ and Cu$_{38}$ was studied in \cite{varsize1} via multi-temperature MD simulations.
Recently, a Markov Chain Monte Carlo algorithm named Grand and Semigrand Canonical Basin Hopping 
allowing additions and removals of atoms was introduced and  used for predicting particularly stable configurations in
multicomponent nanoalloys \cite{varsize2}.

Contrary to the earlier works on clusters with variable {numbers} of particles \cite{varsize1,varsize2}, 
we want to investigate the aggregation process of Lennard-Jones atoms in a more detailed and exhaustive manner starting from 
$N=6$ atoms, as this is the smallest number that admits more then one local energy minimum.
We choose to go along with the network-based approach due to its high level of detailization combined with simplicity and visuality.
Contrary to \cite{varsize1,varsize2}, our approach is completely deterministic. 
First, using deterministic computational techniques, we build a network (a continuous-time Markov chain) 
representing aggregation and dynamics 
of Lennard-Jones clusters. Then we
analyze this network by deterministic methods. Note that deterministic methods, 
whenever their application is feasible, are typically more accurate and 
more efficient than Monte Carlo approaches, whose statistical error decays as $n^{-1/2}$ with the number of samples $n$.
      
Since, to the best of our knowledge, this is the first work that builds a complete network representing an aggregation process and analyzes it,
we start with a very simple aggregation model characterized by the following features.
First, the temperature (the mean kinetic energy of atoms in the cluster) is  maintained constant
throughout the aggregation process. 
Second, new atoms join the cluster one {at} a time arriving at a given fixed stochastic rate.
Third, atoms, once they have joined the cluster, are not allowed to leave it. This assumption
is reasonable provided that the temperature is low enough to render dissociations 
extremely unlikely. 

Our analysis shows that even this simple aggregation model gives results consistent with experimental 
findings, that small clusters tend to have icosahedral packing and form complete icosahedra when {they are} admissible. 
Both processes, attachment and relaxation, taking place in
our aggregation model promote icosahedral packing.
The examination of this simple model gives a reference point for 
further studies of more complicated network models of aggregation processes that will be conducted in our future work.
}

\subsection{A brief summary of main results}
Thus, our goal is
to build a network representing the aggregation and dynamics of Lennard-Jones clusters and analyze it.

We have created such a network for up to 14 atoms. { Our dataset is available at \cite{mydata}.}
The vertices of this network represent local energy minima for each $N$-atom cluster, $6\le N\le 14$.
Energy minima that can be obtained one from another by translations, 
orthogonal transformations, or permutations of atoms are
mapped onto the same vertex. 
For {the sake of} brevity, we will
denote both the $N$-atom Lennard-Jones cluster and the network representing its energy landscape by LJ$_N$.
In each LJ$_{N}$, the local minima are ordered in increasing order of their potential energies. 
The $i$th lowest minimum and the corresponding state of the LJ$_N$ network will {be} denoted by M$N(i)$.
The energy landscapes of LJ$_2$, LJ$_3$, LJ$_4$ and LJ$_5$ are trivial
as they consist of unique potential energy minima: dimer, triangle, tetrahedron, and trigonal bipyramid (bi-tetrahedron)
respectively.
We computed the LJ$_N$ networks for $N=6,\ldots,14$ 
(LJ$_{13}$ is also available at \cite{web}) and connected them 
by evaluating transition probabilities from each vertex of LJ$_N$ to each vertex of LJ$_{N+1}$.
The attachment times are assumed to be exponentially distributed random variables with the parameter $\mu$, 
{so the transition rate along each directed edge (a.k.a. arc) from LJ$_N$ to 
LJ$_{N+1}$ is given by that edge's transition probability multiplied by $\mu$
}.
We did not include arcs from LJ$_{N+1}$ to LJ$_N$, as the transition rates along them would be rather small 
in the considered isothermal aggregation, due to the necessity to 
break at least 3 bonds in order to remove an atom from a cluster. 

Therefore, the resulting  \emph{aggregation/deformation} LJ$_{6-14}$ \emph{network} contains two kinds of edges:
undirected edges connecting vertices {within} each LJ$_N$, and 
directed edges (a.k.a. arcs) connecting LJ$_{N}$ to LJ$_{N+1}$. 
The LJ$_{6-14}$ network is not only \emph{time-irreversible} but also \emph{reducible}. Its invariant probability distribution is supported only on LJ$_{14}$.
We are interested in its transient dynamics. We pose the following question. 
If the aggregation process starts at M6(2), the bicapped tetrahedron local minimum of LJ$_6$, 
{formed} as a result of the attachment of an additional atom to the only minimum of LJ$_5$,
what configurations are most likely to be observed in each LJ$_N$ as the aggregation process proceeds to LJ$_{14}$?

Time-reversibility and/or irreducibility were typically assumed in deterministic methods used for analysis of Lennard-Jones networks,
e.g., the transition path theory tools \cite{cve} need strictly positive invariant distribution to evaluate reactive currents, 
while the eigencurrents are defined so far only for time-reversible and irreducible networks \cite{C-eigen,CG}.
{ Since these standard assumptions do not hold for the LJ$_{6-14}$ network, we have developed new analysis tools.}
In this work, we introduce so-called expected initial and pre-attachment distributions 
 to analyze the aggregation/deformation LJ$_{6-14}$ network. 
 Both of these distributions depend on the attachment rate $\mu$. 
Assuming that an initial probability distribution for LJ$_N$ is given, 
the expected pre-attachment distribution is calculated  from it as the expected probability distribution 
at the attachment time.
Having
 found the expected pre-attachment distribution for LJ$_N$, one can convert it 
 to the expected initial probability distribution for LJ$_{N+1}$
 using the found transition probabilities along the arcs connecting LJ$_N$ and LJ$_{N+1}$. 
 Continuing this process, one can compute the whole sequence of the expected  initial and 
 pre-attachment distributions up to $N=14$ and answer the posed question.
 The inspection of the computed distributions shows 
 at which stage of the process configurations based on icosahedral packing start to dominate.
In particular, the 13-atom icosahedron 
 is the most likely configuration 
 to observe for the 13-atom cluster for a wide range of attachment rates, from low to medium.
 Unsurprisingly, the capped icosahedron, the global minimum of LJ$_{14}$, dominates the initial and the pre-attachment distributions for LJ$_{14}$.
 The computed expected initial and pre-attachment distributions are compared to  the invariant  distributions
 {for the networks LJ$_N$ of fixed cluster size}
 by measuring the normalized root-mean-square discrepancies
 introduced in this work. 
 
 The dominance of local minima based on icosahedral packing is evident from our results for $10\le N\le 14$.
 In order to understand the origin of icosahedral clusters, we examine the attachment process and 
 conduct a structural analysis of  local energy minima for all cluster sizes. 
 The attachment of new atoms converts significant  fractions of local minima of LJ$_N$ 
 with non-icosahedral packing to local minima of LJ$_{N+1}$ with icosahedral packing for $11\le N\le 13$.
 Our results indicate that both processes taking place in the LJ$_{6-14}$ network,  attachment and  relaxation,
 work in favor of the formation of configurations with icosahedral packing.  
 
The rest of the paper is organized as follows.
In Section \ref{sec:construction}, we explain how the LJ$_{6-14}$ network is computed. 
Section \ref{sec:analysis} is devoted to the analysis of the LJ$_{6-14}$ network.
We discuss some perspectives {on} the introduced approach for modeling aggregation processes of interacting particles in Section \ref{sec:conclusion}.

\section{Construction of the Aggregation/Deformation LJ$_{6-14}$ network}
\label{sec:construction}
The LJ$_{6-14}$ network consists of nine LJ$_{N}$ sub-networks, $N=6,\ldots,14$,
connected by arcs representing the attachment of new atoms. 
{The matlab codes developed for building the LJ$_{6-14}$ network are available
in \cite{mydata}.}
 
\subsection{Construction of LJ$_N$ {sub-}networks}
\label{subsec:ljn}
LJ$_6$ has two energy minima separated by a transition state, a.k.a. a Morse-index one saddle 
(the Morse index is the number of negative eigenvalues {of} the Hessian matrix).
The octahedron, the global minimum M6(1) of LJ$_6$, can be formed only due to the structural transition
in  LJ$_6$, while the minimum M6(2), the bicapped tetrahedron, can also arise from the only minimum of LJ$_5$, the
trigonal bipyramid, as a result of the attachment of a new atom.
The LJ$_7$ network containing 4 vertices corresponding to the local minima and 
6 transition states separating distinct vertices was presented in \cite{LJ7}.
We used it as a checkpoint for our techniques.
Since the networks LJ$_{N}$ for $N\le 14$ are relatively small, we aimed at finding the whole set of local minima for each of them.
The found global minima were compared with the list in \cite{wales110}. 
The set of minima of LJ$_{13}$ was taken from \cite{web}\footnotemark[1]. 
The rest of LJ$_N$, $6\le N\le 12$ and $N=14$, had to be 
generated. 
\footnotetext[1]{The dataset for LJ$_{13}$ found in \cite{web} containing 28970 Morse-index one saddles 
significantly oversamples the set of transition states in comparison 
with our networks LJ$_N$, $N\ge 8$. 
Therefore, we computed the set of transition states for LJ$_{13}$ using our technique, 
so that it is sampled consistently with our networks LJ$_{N}$, $6\le N\le 12$ and $N=14$.}

The networks LJ$_N$ for $N\ge 8$ were generated sequentially as follows.
The fast and robust trust region BFGS method \cite{nocedal} was chosen for numerical minimization.
An initial set of local minima was found by $\le10^4$ minimization runs starting from random initial configurations (code {\verb|find_minima.m|} in \cite{mydata}).
Some more local minima were found by $10^3$ hops of the basin hopping method \cite{wales110} starting from 
each initially found minimum (code {\verb|find_minima.m|} in \cite{mydata}). More local minima were found 
as a result of the evaluation of transition probabilities from the minima of LJ$_{N-1}$ to those of LJ$_N$ (see Section \ref{sec:attachment}).
Finally, a few extra local minima were found as a result of our search for transition states starting from each available local minimum
on the other side of the detected Morse index one saddle.

The search for transition states in each LJ$_N$ was accomplished using the technique 
proposed by S. Sousa Castellanos\footnotemark[2] that combined two methods, 
the min-mode method (following the eigenvector associated with the smallest eigenvalue) (e.g. \cite{xiang1,xiang2})
and the shrinking dimer method {\cite{munro,dimer,du-zhang}}, in two {\tt for}-loops going over certain sequences of values of two 
key parameters, the step size in the min-mode method and the threshold value of the eigenvalue 
at which the min-mode method switches to the shrinking dimer (code {\verb|find_saddles.m|} in \cite{mydata}). 
This technique (we named it ``the saddle hunt") has important advantages. 
First, the search for Morse-index one saddles starts at local minima. Hence, the problems of finding 
an initial approximation as in the shrinking dimer method,
or aligning the endpoints as in the string \cite{string1,string2} and the nudged elastic band \cite{neb} methods, 
are eliminated, and the number of runs is equal to the number of local minima.
Second, the saddle hunt method finds collections of distinct Morse-index one saddles starting
from the same local minimum thanks to its {\tt for}-loops. 
In summary, the saddle hunt turned out to be a quite powerful technique. 
It will be reported in details separately.

\footnotetext[2]{S. Sousa Castellanos (East Carolina University) was M. Cameron's MAPS-REU student in Summer 2016}

If two local minima $i$ and $j$ in LJ$_N$ are separated by a transition state $s$, 
the transition rate from  $i$ to $j$ via $s$ is given by Langer's formula \cite{langer}
upgraded to take the point group orders into account as in \cite{wales0}:
\begin{equation}
\label{eq:rates}
L_{i\rightarrow j}^{s} \approx \frac{O_i}{O_{s}} \frac{\left|\lambda_{s}\right|}{2\pi} \sqrt{\frac{\det H_i}{\left|\det H_{s}\right|}} e^{-\beta(V_{s}-V_i)}
\end{equation}
where 
{$\beta^{-1}\equiv k_BT$ is our measure of temperature,}
$\lambda_s$ is the only negative eigenvalue of the Hessian matrix at the saddle $s$, 
$V_i$ and $V_{s}$ are the potential energy values,
$H_i$ and $H_{s}$ are the Hessian matrices, and 
$O_i$ and $O_{s}$ are the point group orders, at $i$ and $s$ respectively.
Note that the corresponding vertices $i$ and $j$ might be separated by multiple edges corresponding to different transition states.
The total transition rate from the state $i$ to the state $j$ is the sum of the transition rates along all edges connecting $i$ and $j$.

We developed a code that computes the point group orders in Eq. \eqref{eq:rates} 
for any {finite} set of points {$X$} in 3D (code {\verb|point_group_order.m|} in \cite{mydata}).
{ Since the center of mass remains invariant for every transformation mapping $X$ onto itself,
we start with grouping the points according to their distances to the center of mass. 
Clearly, if $X$ is mapped onto itself, each subset of points of $X$ equidistant from its center of mass is mapped into itself.
Then} the code makes use of the orbit-stabilizer theorem: given a 
group $G$ acting on a set $X$, for any $i \in X$,
$\left|G\right| = \left|\mathrm{orb}(i)\right| \left|\mathrm{stab}(i)\right|$
where $\mathrm{orb}(i):=\{j\in X~|~gi = j~\text{for some}~g\in G\} $ is the \textit{orbit} of $i$, 
and $\mathrm{stab}(i): = \{g\in G~|~ gi = i\}$ is the \textit{stabilizer} of $i$.
Let $X$ be the set of coordinates $\{\mathbf{r}_k:=(x_k,y_k,z_k)\}_{k=1}^N$ of atomic centers in the cluster,
and $G$ be the point group that we want to find. 
We choose one point $\mathbf{r}_p:=(x_p,y_p,z_p)\in X$ and first  exhaustively 
test all possible orthogonal transformations that leave both $\mathbf{r}_p$ and the center of mass in place, 
while map $X$ onto itself.
{ (Since $X$ is finite, and $G$ is (unless all points in 
$X$ are coplanar) a permutation group of $X$, there are finite 
permutations to test. The number of permutations to test is further 
limited by grouping the points according to their distances to the 
center of mass, as noted above.)
}
Then we exhaustively test whether the atom at $\mathbf{r}_p$ can be mapped to each other atom at 
equal distance from the center of mass, while $X$ is mapped onto itself. 
During both tests, we count all such transformations which 
map $X$ onto itself, and {we} thus obtain $|{\rm stab}(\mathbf{r}_p)|$ and $|{\rm orb}(\mathbf{r}_p)|$. 

Each LJ$_N$ network is time-reversible and irreducible.
With the transition rates given by Eq. \eqref{eq:rates}, its invariant distribution is a row vector $\pi^N$
whose components are given by
\begin{equation}
\label{eq:pi}
\pi^N_i = \frac{ e^{-\beta V_i} O_i^{-1}(\det H_i)^{-1/2}}{ {\sum_j}  e^{-\beta V_j}O_j^{-1}(\det H_j)^{-1/2}},
\end{equation} 
where the sum in the denominator is taken over the whole set of vertices in LJ$_N$. 

\begin{figure}[htbp]
  \includegraphics[width=\textwidth]{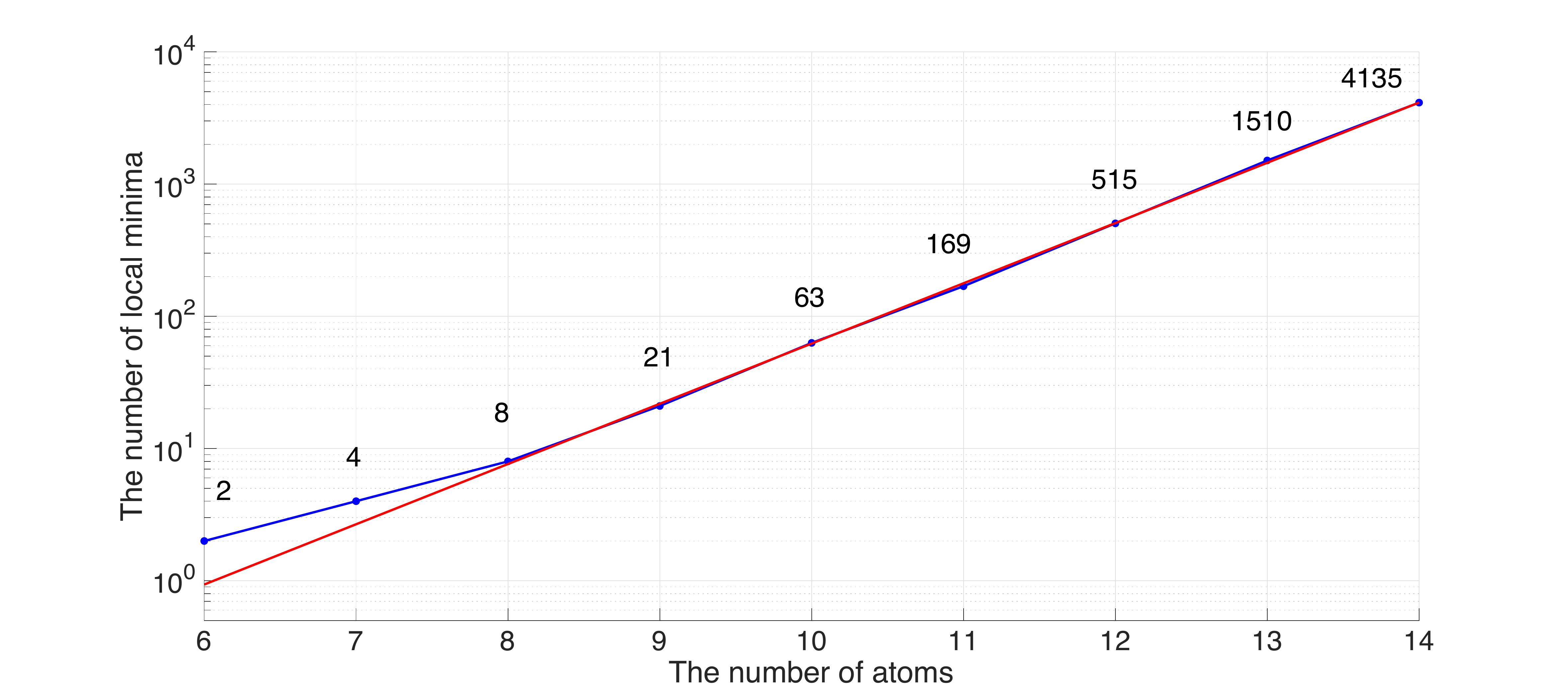}
\caption{
Blue dots: the numbers of local minima in LJ$_N$ for $6\le N\le 14$.
Red line: the least squares fit (ignoring $N=6,7$) given by Eq. \eqref{eq:LJmin}.
}
\label{fig:LJmin}       
\end{figure}

Fig. \ref{fig:LJmin}, showing the numbers of local minima for LJ$_N$, $6\le N\le14$, 
suggests that the number of local minima grows exponentially with the number of atoms.
The least squares fit by an exponential function gives:
\begin{equation}
\label{eq:LJmin}
N_{\min}(N) = 1.7\cdot 10^{-3}\cdot  e^{{1.04}N}.
\end{equation}
{ It was argued in \cite{wallace,stillinger} that the number of 
geometrically different 
{ local} minima of energy landscapes of clusters of particles interacting according to any
short-range pair potential should grow exponentially with the number of particles $N$. 
The estimate of exponential coefficient 0.8 derived in \cite{wallace} 
is in reasonable agreement with our empirical coefficient $1.04$ in Eq. \eqref{eq:LJmin} for small Lennard-Jones clusters.}


\subsection{Connecting LJ$_N$ and LJ$_{N+1}$}
\label{sec:attachment}
A new atom joining a cluster of $N$ atoms configured near a local energy minimum $i$ of LJ$_N$
will cause the cluster to relax to {a neighborhood} of some local minimum $j$ of LJ$_{N+1}$ 
depending on the mutual arrangement of the $N$-cluster and the new atom.
We need to compute the transition probabilities $\gamma^{N\rightarrow N+1}_{ij}$ 
that a minimum $i$ of LJ$_N$ will transform into a minimum $j$ of LJ$_{N+1}$.
Naturally, $\sum_j\gamma^{N\rightarrow N+1}_{ij}= 1$.

\begin{figure}[htbp]
  \includegraphics[width=0.5\textwidth]{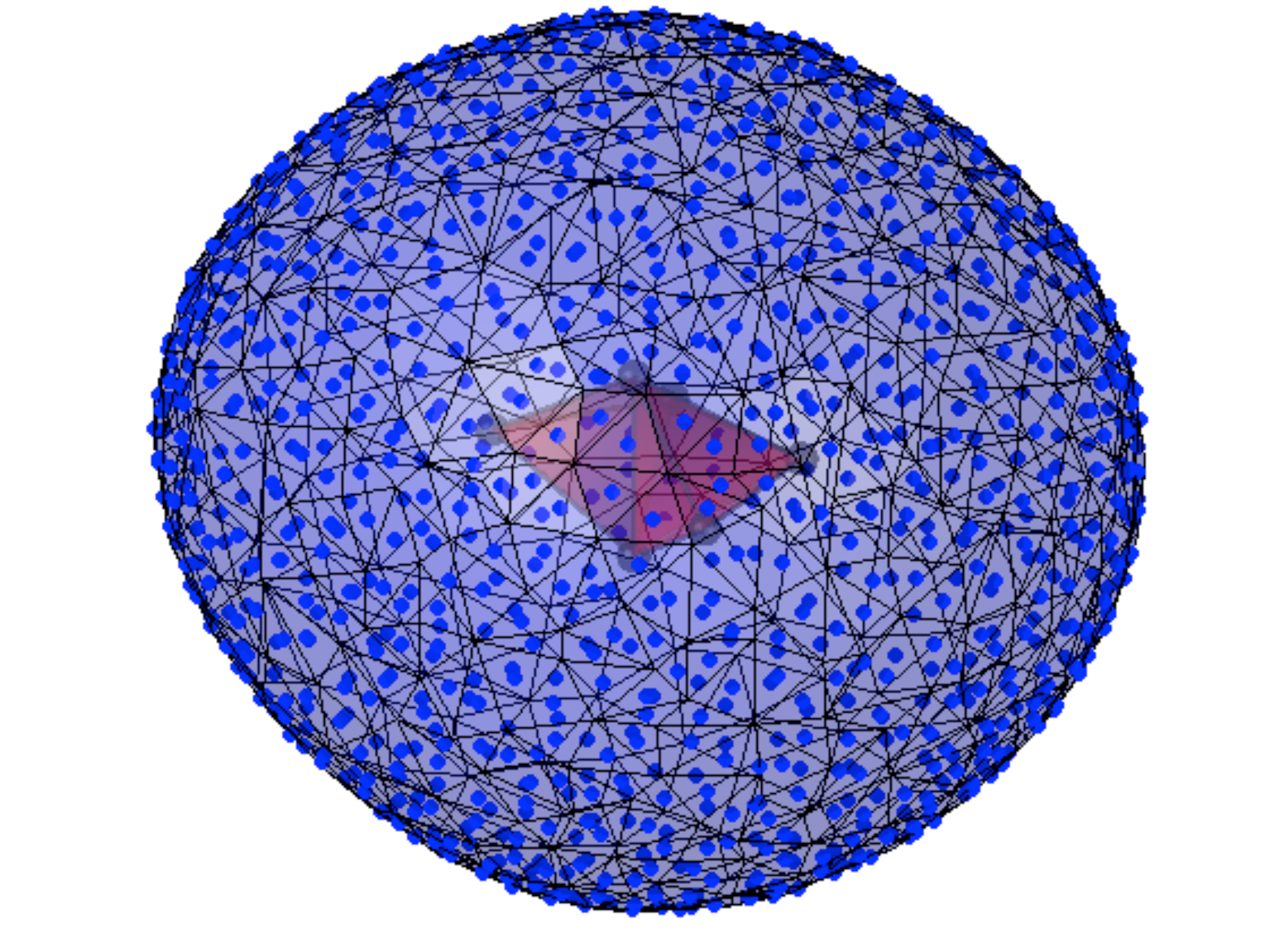}
\caption{The equipotential surface given by Eq. \eqref{eq:isosurface} with $U_0=-0.1$ 
surrounding the bicapped tetrahedron local minimum of LJ$_{6}$.
The surface is triangulated into $1000$ faces. 
Blue dots indicate the centers of the faces.}
\label{fig:isosurface}       
\end{figure}

We propose the following method for estimating the transition probabilities (code \verb|glue_networks.m| in \cite{mydata}).
Let  $U(\mathbf{r})$, $\mathbf{r} = (x,y,z)$, be the potential energy of interaction of 
a new atom at the location $\mathbf{r}$ and the $N$-atom cluster whose atoms are at fixed locations $\{\mathbf{r_k}\}_{k=1}^N$:
\begin{equation}
\label{eq:iso}
U(\mathbf{r}) : = 4 \sum_{k=1}^N \left(|\mathbf{r}-\mathbf{r}_{k}|^{-12} -  |\mathbf{r}-\mathbf{r}_{k}|^{-6}\right).
\end{equation}
Note that $U(\mathbf{r})\rightarrow 0$ as $|\mathbf{r}|\rightarrow \infty$.
Consider the equipotential surface
\begin{equation}
\label{eq:isosurface}
\Sigma:=\{\mathbf{r}\in\mathbb{R}^3~|~U(\mathbf{r}) = U_0,~\min_{1\le k\le N}|\mathbf{r}-\mathbf{r}_{k}| > 2^{1/6}\},
\end{equation}
where $U_0$ is a small-in-absolute-value negative number. In our calculations, we set $U_0=-0.1$.
The condition $~\min_{1\le k\le N}|\mathbf{r}-\mathbf{r}_{k}| > 2^{1/6}$
eliminates the  components  of $\{\mathbf{r}\in\mathbb{R}^3~|~U(\mathbf{r}) = U_0\}$ (if any) lying inside $\Sigma$.
An example of such an equipotential surface surrounding the M6(2) minimum of LJ$_{6}$ is shown in Fig. \ref{fig:isosurface}.
We assume that  the landing site of the new atom arriving from the outer space
on the equipotential surface $\Sigma$ is a uniformly distributed random variable.
We surround every local minimum $i$ of LJ$_{N}$ with the equipotential surface $\Sigma$ given by Eq. \eqref{eq:isosurface},
and triangulate it: 
$$
\Sigma = \bigcup_{m=1}^M\sigma_m,~~{\rm where}~\sigma_m\text{'s are  triangular faces}.
$$ 
The target value of $M$ is 1000.
For each face center, we run minimization using the trust region BFGS {method} with a small maximal trust region radius
and identify the minimum $j$ of LJ$_{N+1}$ to which the run converges. 
For $N\le 8$, it suffices to compare the energy value of the found minimum with the energies of the minima of LJ$_{N+1}$.
For larger $N$, if the energy of the found minimum coincides with that of a minimum $j$ of LJ$_{N+1}$ up to the prescribed tolerance, 
we look for an orthogonal transformation that aligns the found minimum with the minimum $j$.
The transition probability $\gamma^{N\rightarrow N+1}_{ij}$ from  minimum $i$ of LJ$_N$ to  minimum $j$ of LJ$_{N+1}$ is estimated using the formula:
\begin{equation}
\label{tr_prob}
\gamma^{N\rightarrow N+1}_{ij} = \frac{\sum_{m=1}^MA(\sigma_m)\delta_{ij}(m)}{A(\Sigma)},
\end{equation}
{ where $A(\sigma_m)$ is the area of the triangular element $\sigma_m$, $A(\Sigma)$ is the area of the surface $\Sigma$,}
and $\delta_{ij}(m)=1$ if and only if the minimization run for minimum $i$ and face 
$m$ converges to minimum $j$, and $\delta_{ij}(m)=0$ otherwise.

Thus, we connect every pair of vertices $i$ in LJ$_N$ and $j$ in LJ$_{N+1}$
with the arc $(i\rightarrow j)$ whenever $\gamma^{N\rightarrow N+1}_{ij}>0$. Given the attachment rate $\mu$,
the transition rate along the arc $(i \rightarrow j)$ is  $\mu \gamma^{N\rightarrow N+1}_{ij}$.

Such a connection of LJ$_N$ and LJ$_{N+1}$ renders the resulting Markov chain  time-irreversible and reducible. 
Each LJ$_{N}$ component except for the last one created LJ$_{14}$ is transient,
since the attachment causes the process  
to leave each LJ$_{N}$  to LJ$_{N+1}$ without the possibility of return.

\begin{figure}[htbp]
  \includegraphics[width=\textwidth]{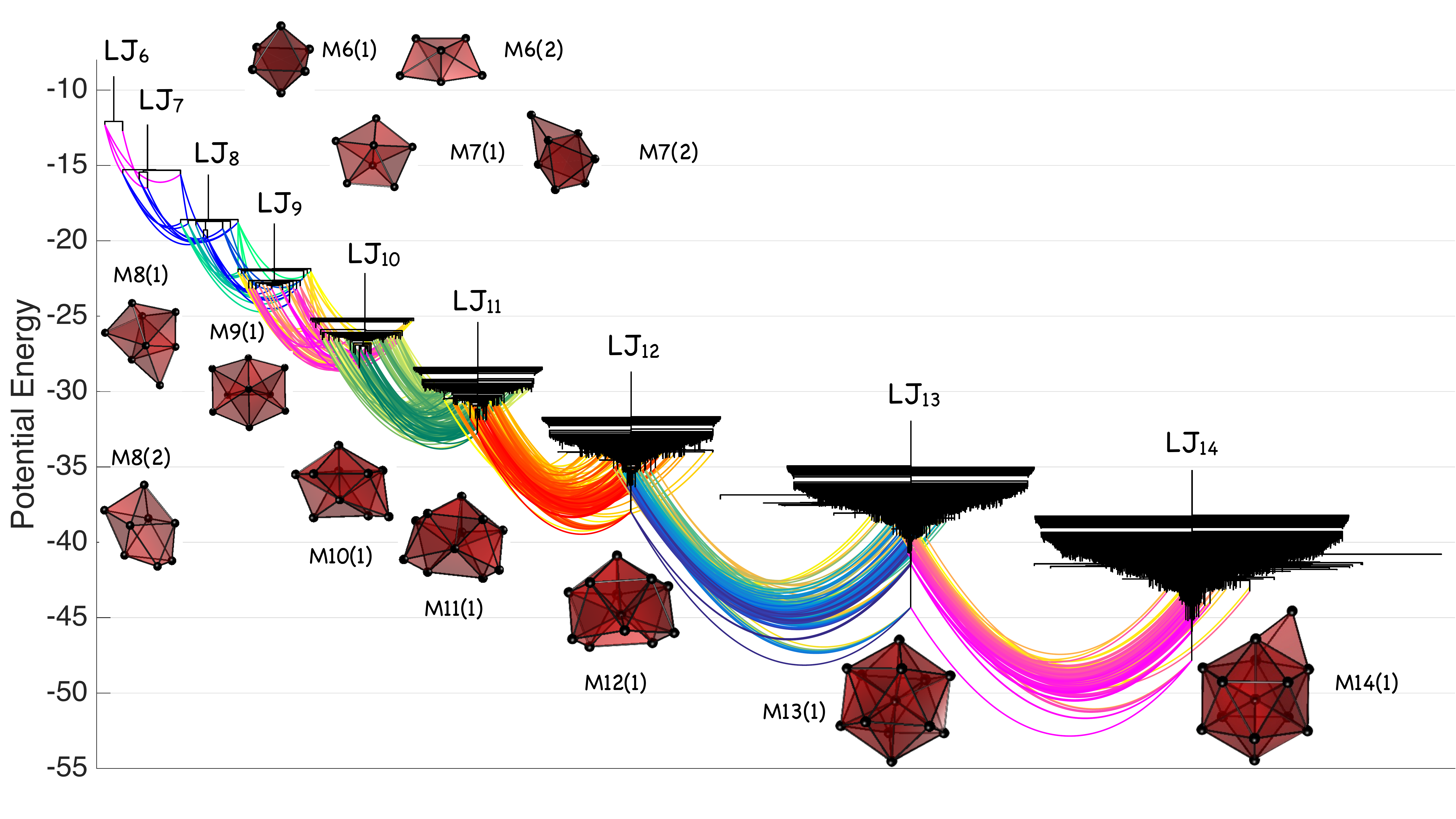}
\caption{
The LJ$_{6-14}$ aggregation/deformation network. The components LJ$_N$
are visualised as disconnectivity graphs, while selected
arcs from LJ$_{N}$ to LJ$_{N+1}$ are depicted with catenary-shaped colored curves.
An arc $(i\rightarrow j)$ from LJ$_N$ to LJ$_{N+1}$ is shown if and only if $i$
is { one of the 50 lowest minima} of LJ$_N$  and $\gamma^{N\rightarrow N+1}_{ij}>0.1$.
For each LJ$_N$, the global minim{um} as well as some local minima are shown.
M$N(n)$  denotes the $n$th lowest minimum of LJ$_N$.  
}
\label{fig:6-14}       
\end{figure}

The created network LJ$_{6-14}$ is visualized in Fig. \ref{fig:6-14}.
Each LJ$_{N}$ is presented as a black disconnectivity graph \cite{becker_karplus}, while selected
arcs from LJ$_{N}$ to LJ$_{N+1}$ are depicted with catenary-shaped colored curves.
An arc $(i\rightarrow j)$ from LJ$_N$ to LJ$_{N+1}$ is shown if and only if $i$
is { one of the 50 lowest minima} of LJ$_N$  and $\gamma^{N\rightarrow N+1}_{ij}>0.1$.
The statistics for the LJ$_{6-14}$ network are presented in Table 1.
\begin{table}[htp]
\caption{The statistics of the aggregation/deformation LJ$_{6-14}$ network.
$N$ is the number of atoms, ``\# min" is the number of minima, 
``\# ts" is the number of found transition states (Morse-index one saddles),
``\# ts, $i\neq j$" is the number of transition states connecting minima mapped to distinct vertices of LJ$_N$,
``$i\neq j$, $\exists$ ts$_{ij}$" is the number of unordered sets of two vertices of LJ$_N$ connected by an edge, i.e.,
half the number of nonzero off-diagonal entries in the generator matrix $L_N$,
``$\langle$degree$\rangle$" is the mean vertex degree of the LJ$_N$ network,
{ ``max degree" is the maximal vertex degree of  the LJ$_N$ network (the  vertex index  (indices) where it is achieved is indicated in the parentheses)},
``\#min0" is the number of vertices $j$ in LJ$_{N}$ such that there is $i$ in LJ$_{N-1}$ such that $\gamma^{{N-1}\rightarrow N}_{ij}>0$.
}
\begin{center}
\begin{tabular}{|c|c|c|c|c|c|c|c|}
\hline
$N$&\# min & \# ts & \# ts, $i\neq j$& $i\neq j$, $\exists$ ts$_{ij}$ &$\langle$degree$\rangle$ & max degree & \# min0\\
\hline
6 & 2 & 3 & 1 & 1 & 1 &  1 (1,2) & 1\\
7 & 4 & 10 & 6 & 5 & 3 & 4 (1) &  4\\
8 & 8 & 51 & 30 & 16 & 7.5 & 18 (1) & 8\\
9 & 21 & 61 & 56 & 45 & 5.33 & 16 (5) & 15\\
10 & 63 & 938 & 700 & 372 & 22.2 & 117 (5) & 60\\
11 & 169 & 756 & 722 & 648 & 8.54 & 53  (12) & 165\\
12 & 515 & 1582 & 1525 & 1410 & 6.04 & 152 (1)& 487\\
13 & 1510 & 4660 & 4512 & 4290 & 5.98 & 306 (1) & 1450\\
14 & 4135 & 13049 & 12630 & 11823 & 6.11 & 1822 (1)& 4109\\
\hline
\end{tabular}
\end{center}
\label{table1}
\end{table}%

\section{Analysis of the Aggregation/Deformation LJ$_{6-14}$ network}
\label{sec:analysis}
The LJ$_{6-14}$ aggregation/deformation network is time-irreversible and reducible.
Its states lying in LJ$_N$ for $6\le N\le 13$ are transient. 
In addition, {although} we did not compute the LJ$_{15}$ network, we can 
assume that a new atom attaches to LJ$_{14}$ as happens for LJ$_N$, $N\le 13$, and treat the LJ$_{14}$ component as transient as well.
This is equivalent to adding an additional vertex $v_{15}$ to LJ$_{6-14}$ representing LJ$_{15}$, shooting arcs
from every vertex of LJ$_{14}$ to $v_{15}$, and setting the transition rates along these arcs to $\mu$.
We would like to study the relaxation process in the LJ$_{6-14}$ network 
starting at M6(2), the bicapped tetrahedron local minimum of LJ$_6$, that is obtained from the only minimum of LJ$_5$
by attaching an extra atom.  The proposed analysis approach is described in Section \ref{subsec:analysis}.
Central to it is the calculation of the expected initial and pre-attachment distributions for each LJ$_N$, $6\le N\le 14$.
The results are presented in Section \ref{subsec:results}. 
In Section \ref{sec:distr},  
the obtained expected initial and pre-attachment distributions are compared to  the invariant  distribution {for each LJ$_N$}.
Finally, a structural analysis of local energy minima is conducted in Section \ref{sec:struct}, and the formation mechanism of 
configurations with icosahedral packing is investigated.


\subsection{Analysis method}
\label{subsec:analysis}
For each number of atoms $N$, $6 \leq N \leq 14$, we compute
two probability distributions: the expected probability distribution LJ$_N$ after the attachment of the $N$th atom,
and the expected distribution in LJ$_N$
right before the attachment of the $(N+1)$st atom. 
 We refer to them as \emph{the expected initial and pre-attachment distributions} 
 and denote them by $p^N_0$ and $p^N_e$ respectively. 

The expected initial distribution for LJ$_6$ is $p^6_0=[0,1]$, where $0$ corresponds to the global minimum M6(1), the octahedron,
while $1$ corresponds the bicapped tetrahedron M6(2).

The expected pre-attachment distribution  for  LJ$_N$ can be found as follows. 
We consider   two random variables: the continuous random variable 
$T$, the attachment time, 
i.e., the time between the arrivals of two consecutive new atoms,
and the discrete random variable 
$S$, which indicates the state/vertex immediately before attachment. 
The joint 
probability density $f^N_{S,T}(s,t)$ can be expressed as 
\begin{equation}
\label{eq:jpd}
f_{S,T}^N(s,t) = \mathbb{P}^N(S = s | T = t) f_T(t).
\end{equation}

We assume  that $T$ is an exponentially distributed random variable with  the probability density function $f_T(t) = \mu e^{-\mu t}$.

The probability $\mathbb{P}^N(S = s | T = t)$ can be found from the following considerations.
Suppose that the initial probability distribution in  LJ$_{6-14}$ is supported within LJ$_N$, where $6\le N\le 14$.
Let $p^{N}(t)$ be the subset of components of the probability distribution corresponding to {the set of} states of LJ$_N$.
{ The conditional probability distribution  $\hat{p}^N(t)$ in LJ$_N$ conditioned on the fact that the system remains in LJ$_N$ at time $t$}
is given by
\begin{equation}
\label{eq:nondec}
\hat{p}^N(t) = \sum_{k=0}^{M_N-1} (p^N_0 \phi_N^k) e^{-\lambda_N^k t}(P_N\phi_N^k)^T,\quad 0\le t< T.
\end{equation}
Here $M_N$ is the number of states in LJ$_N$; $p^N_0=p^N(0)$ is the initial distribution; 
$-\lambda^k_N$'s are the eigenvalues of $L_N$, the restriction of the generator matrix of LJ$_{6-14}$ to LJ$_N$;
$\phi_N^k$ and $(P_N\phi_N^k)^T$ are the
corresponding right and left eigenvectors respectively;
$P_N$ is the diagonal matrix with the invariant distribution $\pi_N$ for LJ$_N$
given by Eq. \eqref{eq:pi} along its diagonal. 
The eigenvectors are normalized so that $\Phi_N \Phi_N^TP_N=\Phi_N ^TP_N\Phi_N = I$, 
where $\Phi_N = [\phi^0_N,\ldots,\phi_{N}^{M_N-1}]$ is the matrix whose columns are the right eigenvectors.
Hence, right before the arrival of the new atom at time $t$, 
$\mathbb{P}(S = s | T = t) = \hat{p}_s^N(t)$, the 
$s$th component of $\hat{p}^N(t)$.

Integrating out the attachment time $T$, we obtain the expected probability distribution at the moment 
right before the arrival of {the} $(N+1)$st atom:
\begin{align}
p^N_e(s) \equiv \mathbb{P}^N(S = s) &= \int_0^\infty {f_{S,T}^N(s,t) dt} \notag\\
&= \int_0^\infty {\mathbb{P}^N(S = s | T = t) f_T(t) dt}  \notag\\
&= \int_0^\infty {\hat{p}^N_s(t) \mu e^{-\mu t} dt}  \notag\\
&= \mu \sum_{k=0}^{N-1} (p^N_0 \phi_N^k) \left(\int_0^\infty {e^{-(\mu+\lambda^k_N) t} dt}\right) (P_N \phi_N^k)^T_s \notag\\
&= \sum_{k=0}^{N-1} (p^N_0 \phi_N^k) \left(\frac{\mu}{\mu+\lambda_N^k}\right) (P _N\phi_N^k)^T_s. \label{eq:dec}
\end{align}
Therefore, the expected pre-attachment distribution is given by
\begin{equation}
\label{eq:preat}
p_e^N = \mu p_0^N\Phi_N(\mu I-\Lambda_N)^{-1}  \Phi_N^TP_N =  \mu p_0^N(\mu I - L_N)^{-1},
\end{equation}
where $\Lambda_N = {\rm diag}\{0,-\lambda_N^1,\ldots,-\lambda_N^{M_N-1}\}$.

Once the expected pre-attachment distribution $p_e^N$ is computed using Eq. \eqref{eq:preat}, 
one can obtain the expected initial distribution for LJ$_{N+1}$ by multiplying the
pre-attachment distribution by the $M_N\times M_{N+1}$ transition matrix 
$\Gamma^{N\rightarrow N+1} = (\gamma^{N\rightarrow N+1}_{ij})$, 
where $\gamma^{N\rightarrow N+1}_{ij}$ are given by Eq. \eqref{tr_prob}:
\begin{equation}
\label{eq:exit}
p_0^{N+1} = p_e^N\Gamma^{N\rightarrow N+1}.
\end{equation}

Starting from $p_0^6 = [0,1]$ and using Eqs. \eqref{eq:preat} and \eqref{eq:exit},
one can compute the sequence of the expected pre-attachment and initial distributions
$p_e^N$ and $p_0^{N+1}$ for $6\le N\le 13$ and the pre-attachment distribution $p_e^{14}$.

\subsection{The sequence of the initial and the pre-attachment distributions}
\label{subsec:results}
\begin{figure}[htbp]
\includegraphics[width=\textwidth]{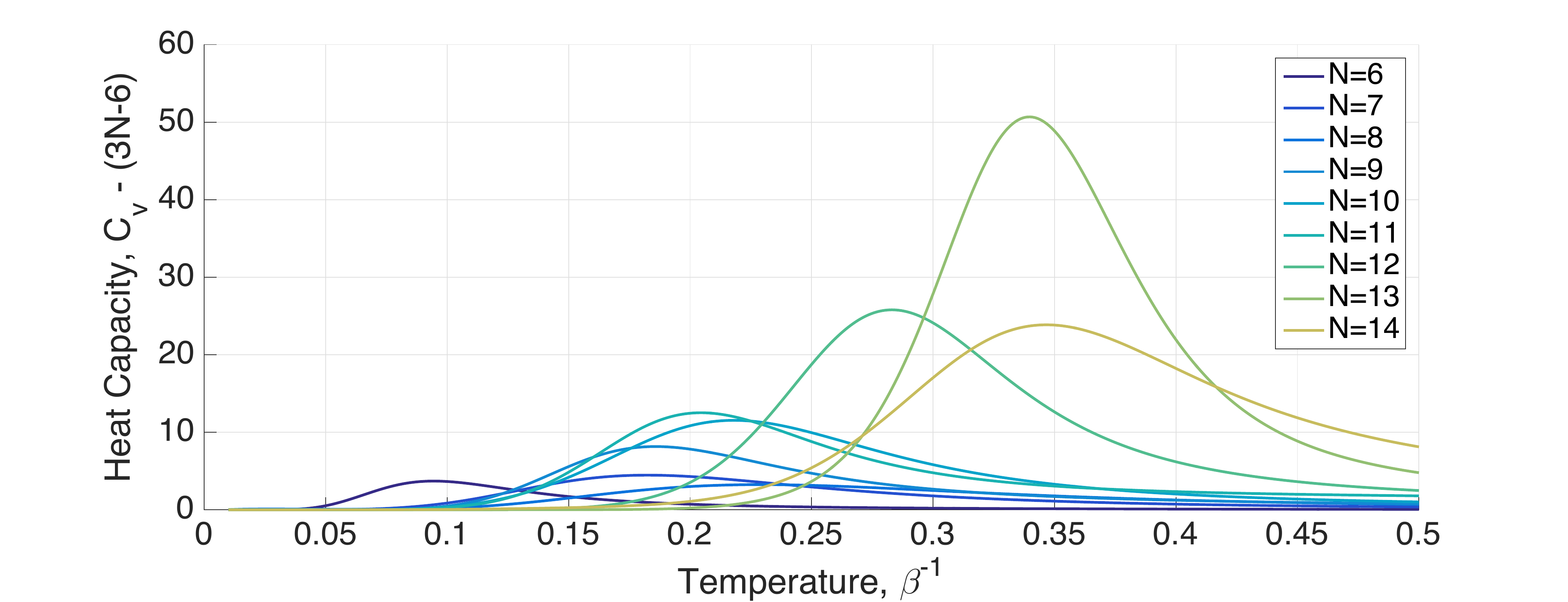}
\caption{The heat capacities $C_v - (3N-6)$
for LJ$_N$ networks, $6\le N\le 14$. 
}
\label{fig:Cv}
\end{figure}

We have calculated 
the expected initial and pre-attachment distributions
for the transient LJ$_N$ networks, $6\le N\le 14$,  for the range of attachment rates
$10^{-4} \leq \mu \leq 10^{4}$ and three values of temperatures:  $k_BT\equiv\beta^{-1}=0.06$, 0.08, and 0.10.
{ As was mentioned in Section \ref{sec:model}, the temperature should be low enough to justify the assumption that
detachments can be neglected. A reasonable criterion for choosing appropriate temperature values is that they 
lie below the maximizer of the heat capacity\cite{wales-thermo} of the cluster LJ$_{N}$  corresponding to the 
major structural (phase) transition. 
In LJ$_N$, $7\le N\le 14$, the single maximum of the heat capacity corresponds to the phase transition 
from solid to liquid-like configurations.
We remind that the heat capacity of a cluster is given by
$$
C_v(\beta^{-1}):=\frac{\partial\langle V\rangle}{\partial \beta^{-1}} = 
\frac{\partial}{\partial \beta^{-1}}\left(\frac{\sum_{i}V_iO_i^{-1}\sqrt{\det H_i}e^{-\beta V_i}}{\sum_{i}O_i^{-1}\sqrt{\det H_i}e^{-\beta V_i}}\right),
$$  
where the sum is taken over all minima $i$ of LJ$_N$.
Fig. \ref{fig:Cv} shows that our chosen temperatures $\beta^{-1}=0.06$, 0.08, and 0.10 
are below the maximizers of the heat capacity $C_v$  for clusters LJ$_N$, $7\le N\le 14$, and around it for {LJ$_6$}.
Note that the only structural transition in LJ$_6$ is the one from the dominance of M6(1), the octahedron, to the dominance of M6(2).
LJ$_6$ is too small to admit liquid-like states.}

\begin{figure}[htbp]
\includegraphics[width=\textwidth]{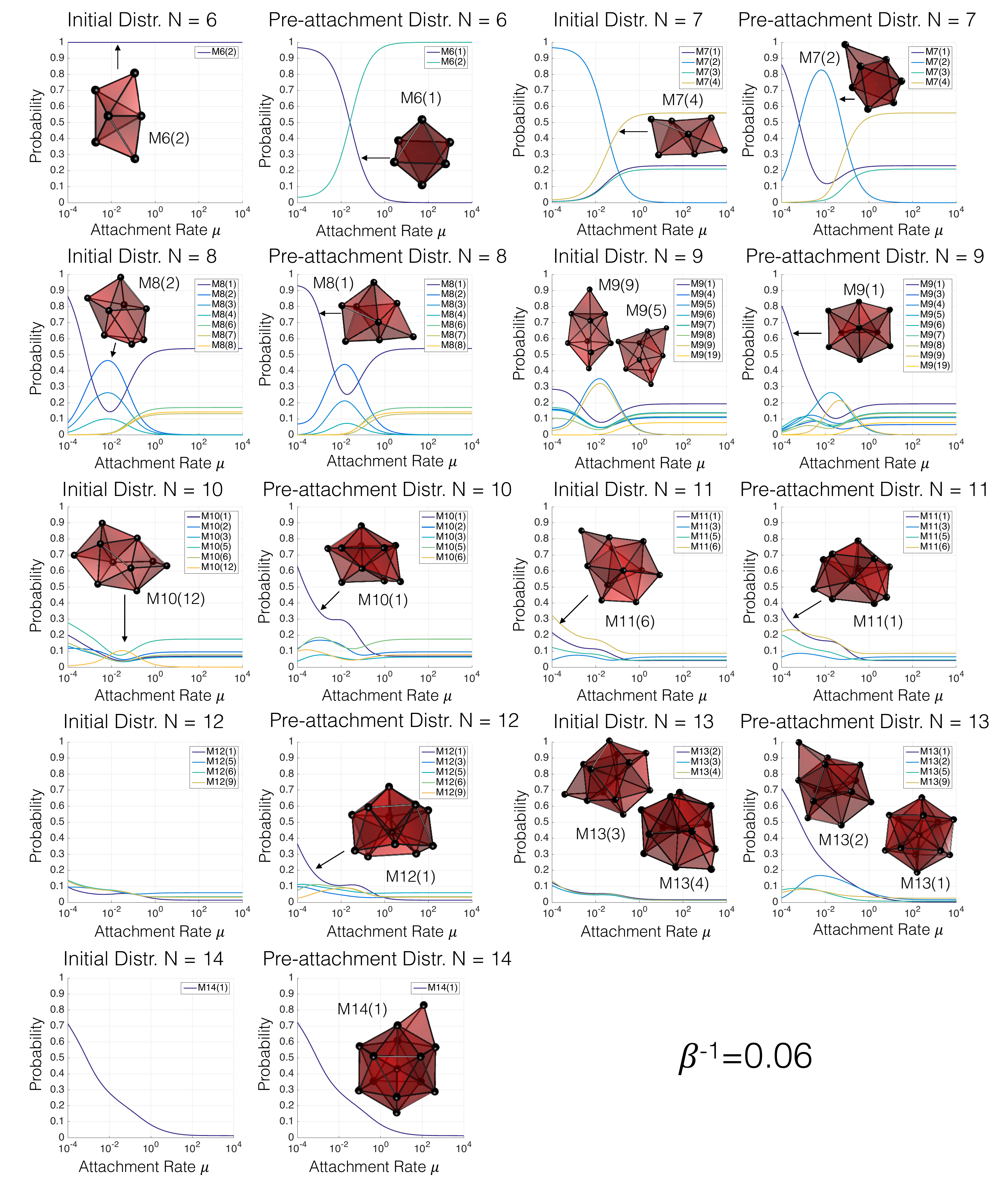}
\caption{The expected initial and pre-attachment distributions 
for the aggregation process from $N=6$ to $N=14$ atoms at $\beta^{-1} = 0.06$.}
\label{fig:distr1}
\end{figure}

\begin{figure}[htbp]
\includegraphics[width=\textwidth]{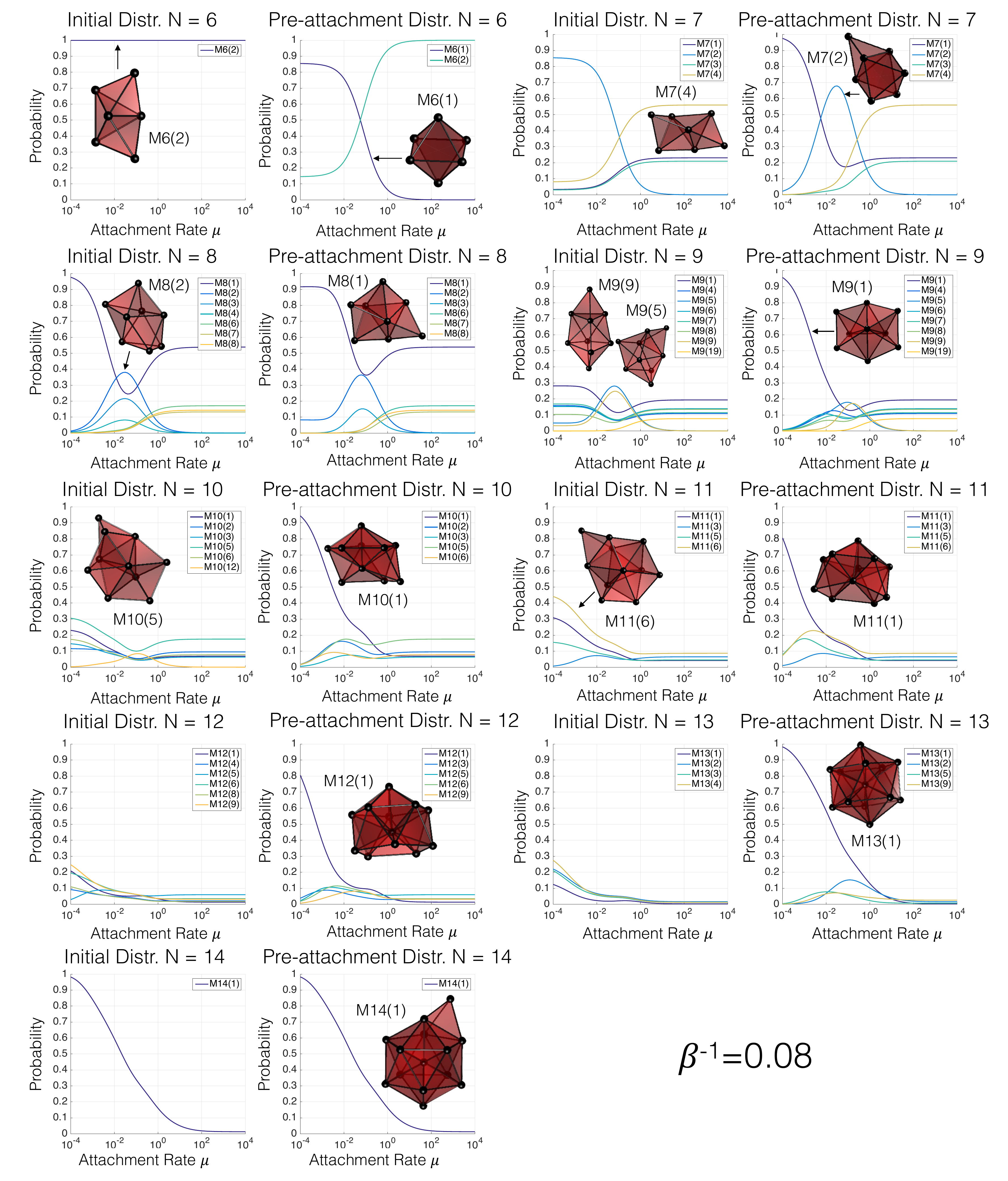}
\caption{The expected initial and pre-attachment distributions 
for the aggregation process from $N=6$ to $N=14$ atoms at $\beta^{-1} = 0.08$.}
\label{fig:distr2}
\end{figure}

\begin{figure}[htbp]
\includegraphics[width=\textwidth]{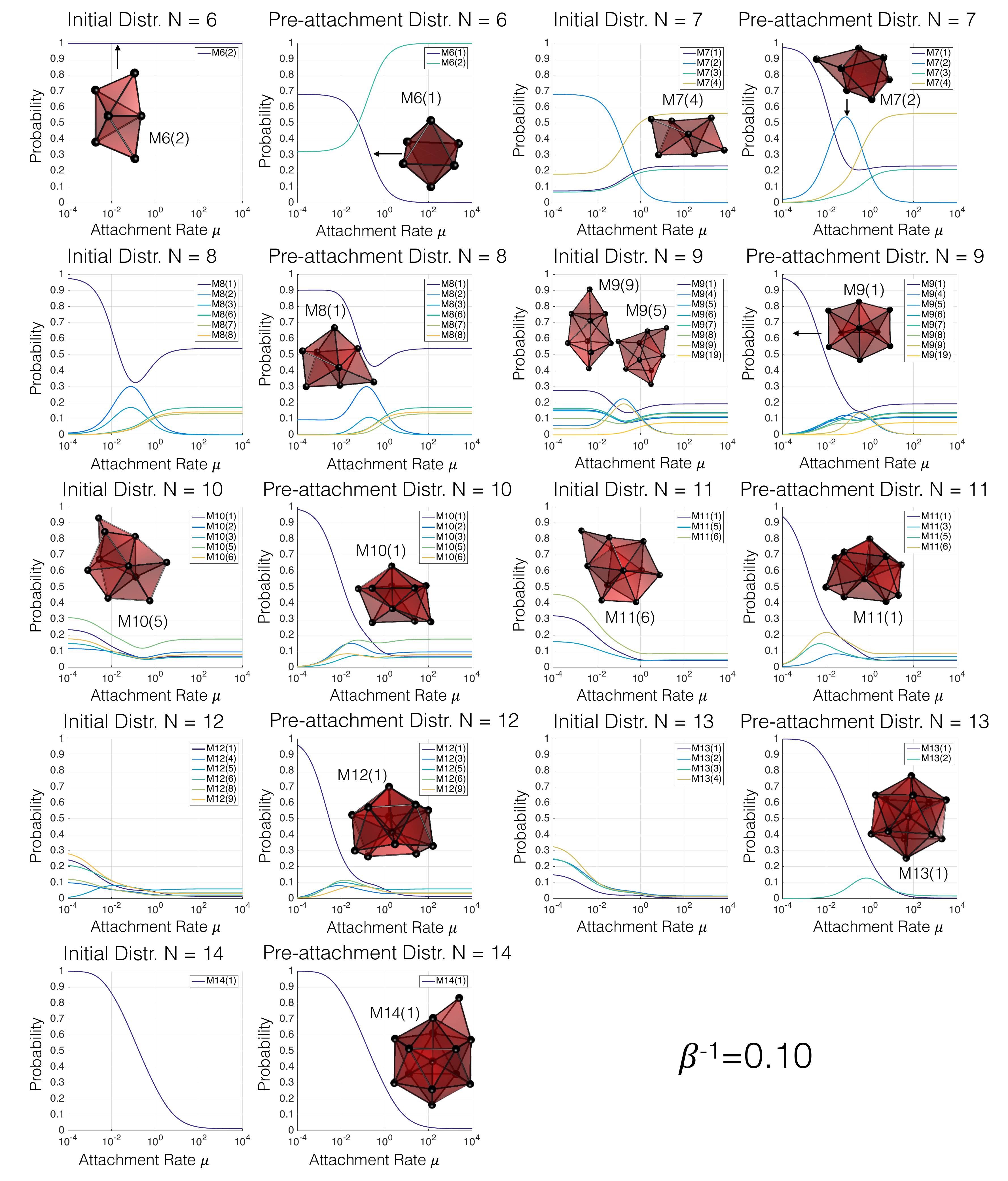}
\caption{The expected initial and pre-attachment distributions 
for the aggregation process from $N=6$ to $N=14$ atoms at $\beta^{-1} = 0.10$.}
\label{fig:distr3}
\end{figure}

The resulting distributions for $10^{-4}\le \mu\le 10^4$ are shown in 
Figs. \ref{fig:distr1}, \ref{fig:distr2}, and \ref{fig:distr3} for $\beta^{-1}=0.06,~0.08$, and $0.10$ respectively.
To avoid cluttering near the $\mu$-axis, only those components of the distributions 
that attain at least 7\% likelihood for some values of 
$\mu$ are shown.

Eq. \eqref{eq:dec} implies that 
the expected pre-attachment distribution for LJ$_N$ approaches 
the invariant distribution as $\mu\rightarrow 0$, and 
approaches the expected initial 
distribution as $\mu\rightarrow\infty$. Indeed,   
the factor ${\mu}(\mu+\lambda^k_N)^{-1}$ tends to zero as $\mu\rightarrow 0$ for all $k\ge 1$, and
tends to 1 as $\mu\rightarrow \infty$ for all $k\ge 0$. 
This is consistent with our results (Figs. \ref{fig:distr1} -\ref{fig:distr3}). For all $6\le N\le 14$, 
the expected pre-attachment distributions for $\mu=10^{-4}$ are nearly the invariant distributions at the corresponding values of $\beta$,
while for $\mu=10^4$, they are nearly the corresponding expected initial 
distributions. 
As the attachment rate $\mu$ becomes large, the relaxation process in each LJ$_N$ cluster {is} limited, resulting in
broad expected initial and pre-attachment distributions
as one can infer from Figs. \ref{fig:distr1}-\ref{fig:distr3}.

\begin{figure}[htbp]
\includegraphics[width=\textwidth]{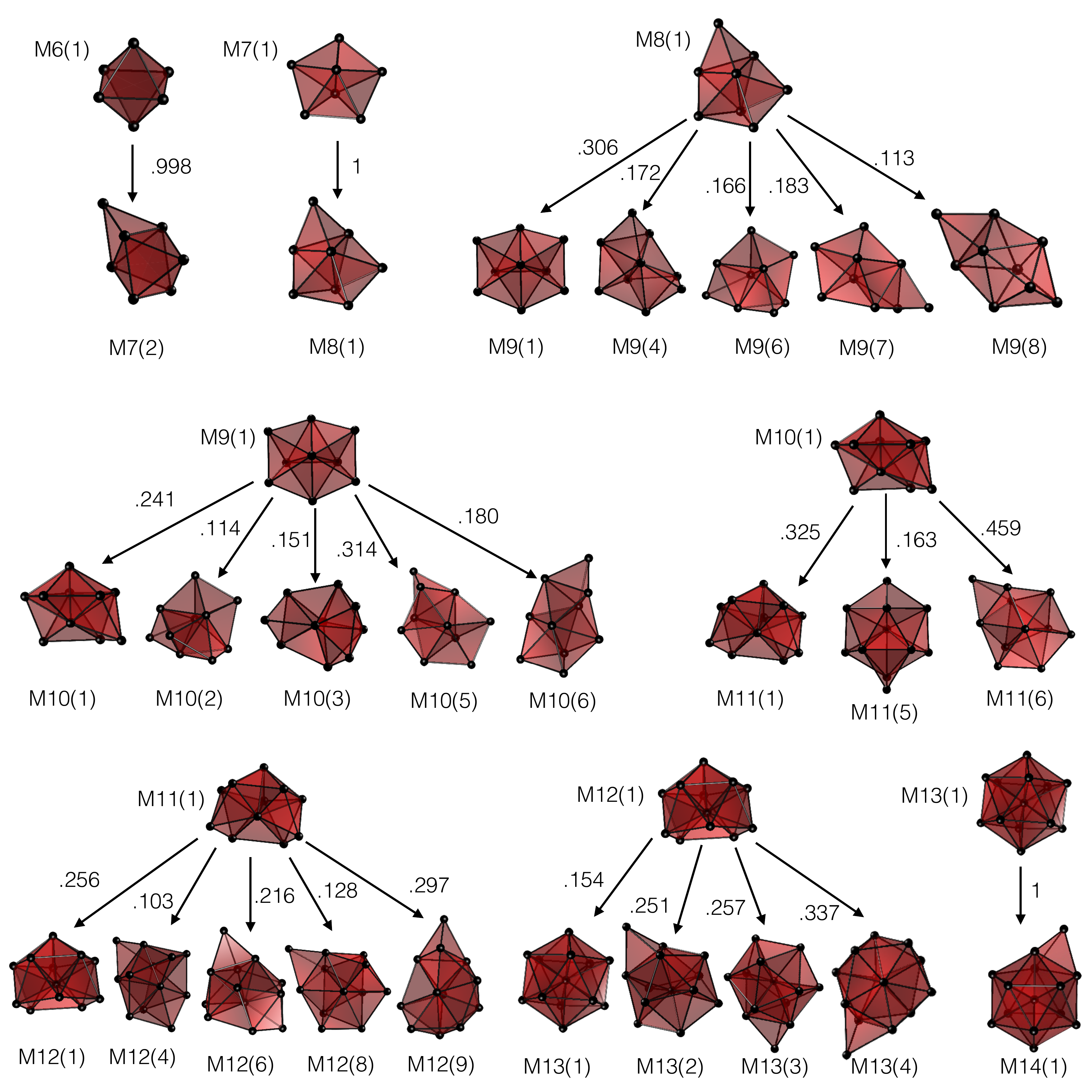}
\caption{The transitions from the global minima of LJ$_N$ to configurations of 
LJ$_{N+1}$, $6\le N\le {13}$,
happening with probabilities at least 0.1. 
The numbers next to the arrows indicate the transition probabilities.}
\label{fig:trans}
\end{figure}

The global minima for $7\le N\le 14$ 
are based on icosahedral packing { (i.e., can be completed to nearly regular icosahedra merely by adding atoms)},
while the one for $N=6$ is the octahedron, which is an elementary cell of a face-centered cubic crystal.
The transitions from the global minima of LJ$_N$ to configurations of LJ$_{N+1}$, $6\le N\le 14$,
happening with probabilities at least 0.1 are illustrated in Fig. \ref{fig:trans}. 

\begin{figure}
\includegraphics[width=\textwidth]{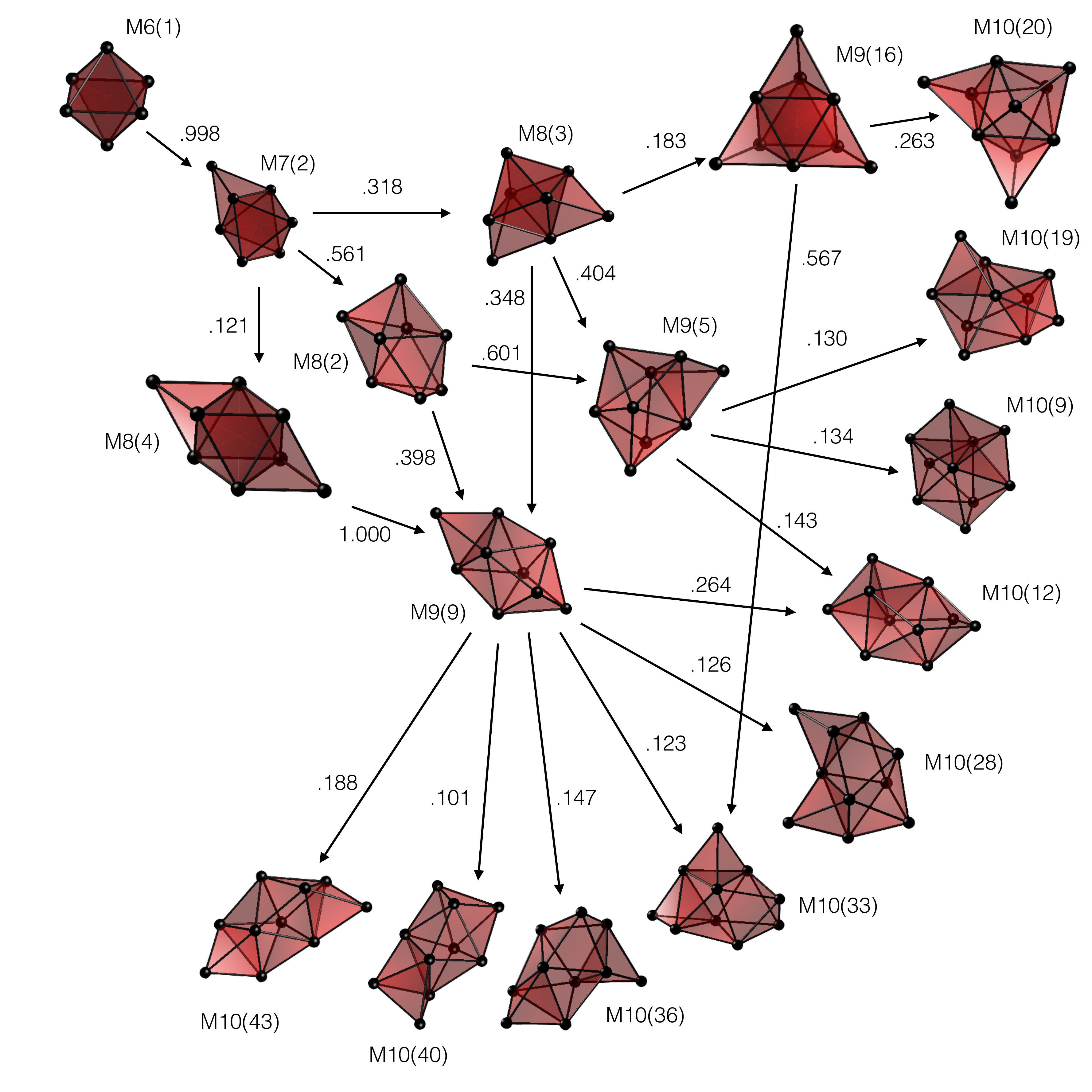}
\caption{The heritage cascade of the global minimum M6(1) of LJ$_6$, 
the octahedron, for up to 10 atoms.
All transition probabilities exceeding 0.1 are displayed. 
The numbers next to the arrows indicate the transition probabilities.}
\label{fig:nonico}
\end{figure}

\begin{figure}[htbp]
\includegraphics[width=\linewidth]{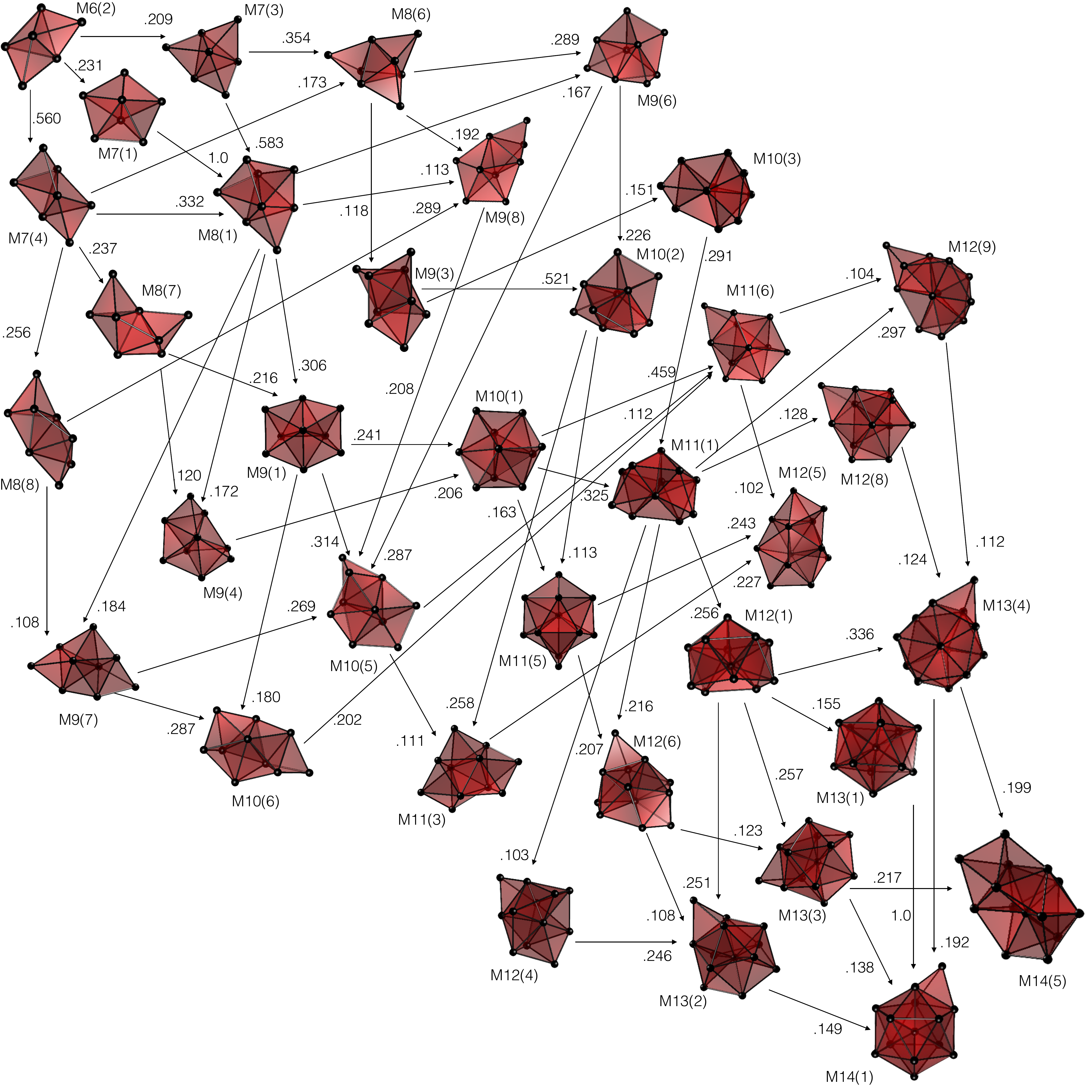}
\caption{The heritage cascade of the  minimum M6(2) of LJ$_6$, the bicapped trigonal bipyramid. 
Only some configurations with probabilities exceeding 0.1 
in the expected initial or pre-attachment distributions in Figs. \ref{fig:distr1}-\ref{fig:distr3} and displayed,
and only transition probabilities exceeding 0.1 are shown. 
The numbers next to the arrows indicate the transition probabilities.} 
\label{fig:ico}
\end{figure}

One can observe two types of persisting clusters in Figs. \ref{fig:distr1} - \ref{fig:distr3}: icosahedral and non-icosahedral.
The probabilities of the heirs of the 6-atom octahedron, 
M7(2), M8(2), M8(3), M9(5), M9(9), and M10(12), {peak in the mid-range of the attachment rate $\mu$ and }are especially prominent for $\beta^{-1}=0.06$ (Fig. \ref{fig:distr1}).
A more complete heritage cascade of non-icosahedral clusters up to $N=10$ is shown in Fig. \ref{fig:nonico}. 
The icosahedral heritage cascade is partially displayed in Fig. \ref{fig:ico} (partially, as it quickly becomes too broad). 
Comparing these cascades, we observe that  
the non-icosahedral one involves only high-energy minima of LJ$_{10}$: the lowest of them is {M10(9)}.
On the contrary, the icosahedral heritage cascade 
 involves all global minima and many other low-energy minima 
 for $7\le N\le 14$. 

\begin{figure}[htbp]
\includegraphics[width=\linewidth]{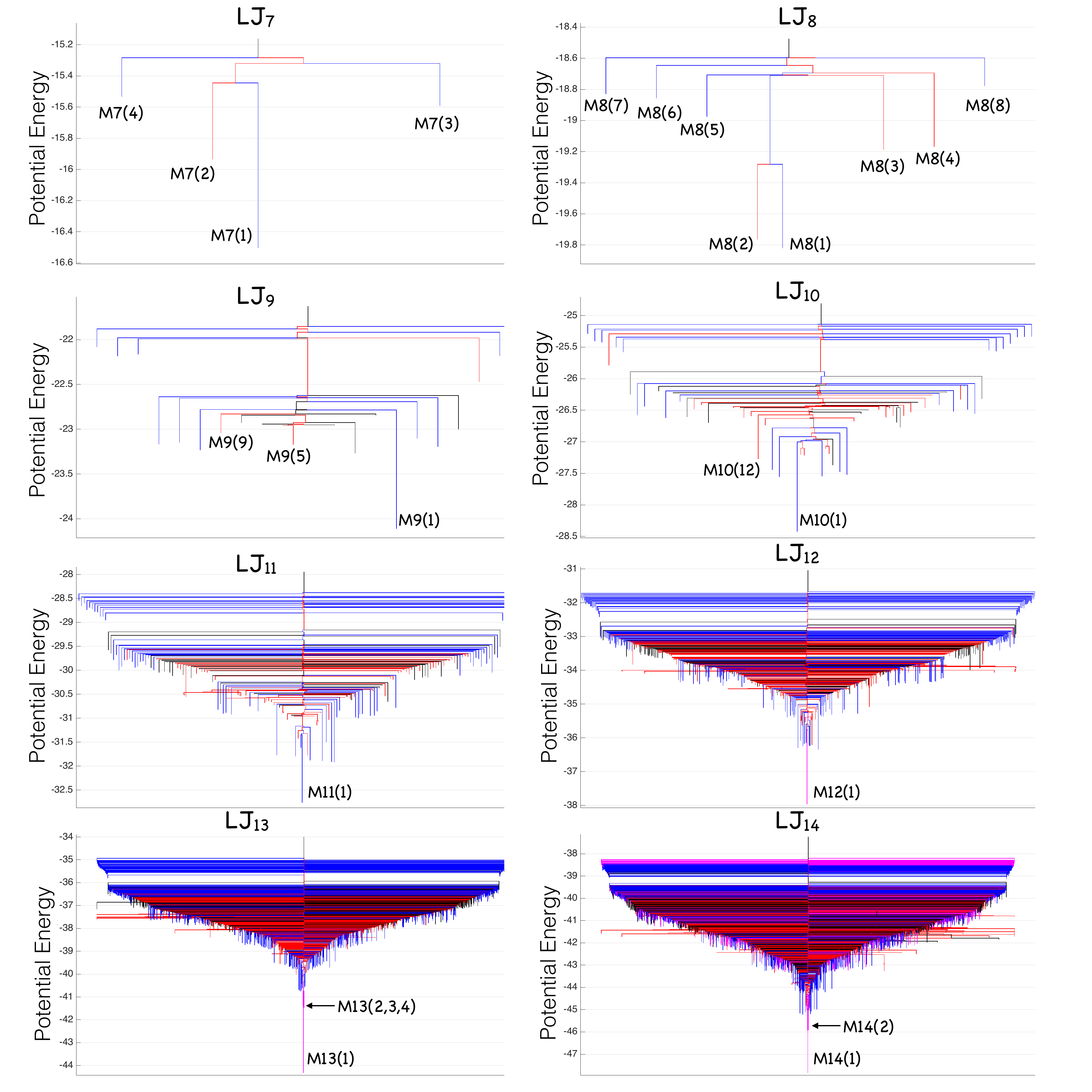}
\caption{The disconnectivity graphs for $7\le N\le 14$. Blue ($b^N(i) > a^N(i)$) and magenta ($b^N(i) < a^N(i)$) {leaves} correspond to minima with icosahedral packing.
Red  ($b^N(i) < a^N(i)$)  and black ($b^N(i) > a^N(i)$) {leaves} correspond to minima with non-icosahedral packing.
} 
\label{fig:redblue}
\end{figure}


The aggregation process involves two kinds of processes: attachment and relaxation.
In order to examine the aggregation process as $\mu\rightarrow \infty$, 
it is instructive to compare two aggregation processes involving only attachment, 
one starting from M6(1) and the other one starting from M6(2).
The corresponding probability distributions for LJ$_{N}$ are given, respectively, by
\begin{equation}
\label{att}
a^{N}:=[1,0]\Gamma^{6\rightarrow 7}\ldots\Gamma^{N-1\rightarrow N}\quad{\rm and}\quad 
b^{N}:=[0,1]\Gamma^{6\rightarrow 7}\ldots\Gamma^{N-1\rightarrow N}.
\end{equation}
Now, for each state in each LJ$_N$ network, $6\le N\le 14$, we compare the distributions $a^N$ and $b^N$.
Fig. \ref{fig:redblue} displays the disconnectivity graphs where the states $i$  are
plotted red/magenta or blue/black depending on whether $a^N(i) > b^N(i)$ or  $a^N(i) < b^N(i)$ respectively. 
It is evident from Fig. \ref{fig:redblue} that the probabilities for the global minima of LJ$_{N}$, $12\le N\le 14$,  
to form starting from M6(1) are larger than those starting from M6(2). 
This is an interesting fact, and we investigate it in more detail.
%

Let $A_N$  and $B_N$ be the subsets of states of LJ$_N$ defined by
\begin{equation}
\label{AB}
A_N:=\{i~|~a^N(i) > b^N(i)\},\quad
B_N:=\{i~|~a^N(i) < b^N(i)\}.
\end{equation}
The numbers of states in $A_N$ and $B_N$ as well as the probabilities to find the LJ$_N$ cluster in  $A_N$ and $B_N$
assuming the invariant distribution in LJ$_N$ for the temperatures $\beta^{-1} = 0.06, 0.08, 0.10$ are shown in Table 2.
The sizes of the sets $B_N$ grow slower than those of $A_N$, and $|A_N|$ surpasses $|B_N|$ at $N=10$.
Meanwhile, the sets $A_N$ contain only low occupancy states for $N=7,8$, and extremely low occupancy (high energy) states
for $N=9,10,11$. However, for $N\ge 12$, the sets $A_N$ acquire the global minima and their probabilities switch to almost one
at the considered temperatures.  
\begin{table}[htp]
\caption{The numbers of states in the sets $A_N$ and $B_N$, $7\le N\le 14$ defined by Eq. \eqref{AB}
and the probabilities to find the N-atom cluster in them assuming the invariant distributions in LJ$_N$.
}
\begin{center}
\begin{tabular}{|c|c|c|}
\hline
$N$, $\beta^{-1}$& $|A_N|$, $\mathbb{P}(A_N)$ & $|B_N|$, $\mathbb{P}(B_N)$\\
\hline
$\mathbf{N = 7}$ &    $|A_7| = 1$ &   $|B_7|=3$\\
\hline
0.06 & 3.797e-4 & 9.996e-1\\
0.08 & 4.072e-3 & 9.959e-1\\
0.10 & 1.667e-2 & 9.833e-1\\
\hline
$\mathbf{N = 8}$ & $|A_8| = 3$ &  $|B_8|=5$\\
\hline
0.06 & 6.721e-2 & 9.328e-1\\
0.08 & 8.372e-3 & 9.163e-1\\
0.10 & 9.610e-2 & 9.039e-1\\
\hline
$\mathbf{N = 9}$ & $|A_9| = 10$ & $|B_9|=11$\\
\hline
0.06 & 2.445e-6 & 1\\
0.08 & 1.679e-4 & 1\\
0.10 & 2.359e-3 & 9.976e-1\\
\hline
$\mathbf{N = 10}$ &  $|A_{10}| = 38$ & $|B_{10}|=25$\\
\hline
0.06 & 1.522e-7 & 1\\
0.08 & 1.789e-5 & 1\\
0.10 & 3.408e-4 & 9.996e-1\\
\hline
$\mathbf{N = 11}$ & $|A_{11}| = 100$ &  $|B_{11}|=69$\\
\hline
0.06 & 1.023e-7 & 1\\
0.08 & 1.458e-5 & 1\\
0.10 & 3.666e-4 & 9.996e-1\\
\hline
$\mathbf{N = 12}$ & $|A_{12}| = 331$ & $|B_{12}|=170$\\
\hline
0.06 & 1 & 3.766e-11\\
0.08 & 1 & 4.781e-8\\
0.10 & 1 & 3.568e-6\\
\hline
$\mathbf{N = 13}$ & $|A_{13}| = 1038$ & $|B_{13}|=472$\\
\hline
0.06 & 1 & 1.802e-23\\
0.08 & 1 & 6.949e-17\\
0.10 & 1 & 6.398e-13\\
\hline
$\mathbf{N = 14}$ &  $|A_{14}| = 2877$ &  $|B_{14}|=1257$\\
\hline
0.06 & 1 & 4.787e-18\\
0.08 & 1 & 3.870e-13\\
0.10 & 1 & 3.514e-10\\
\hline
\end{tabular}
\end{center}
\label{table2}
\end{table}%

\subsection{Comparison to Invariant Distributions}
\label{sec:distr}
In this Section, we introduce the normalized root-mean-square (NRMS) deviation 
and use it to compare the computed expected initial and pre-attachment distributions to the invariant distribution for each LJ$_N$.

Let $\pi$
be a probability distribution. 
The  most different from $\pi$ is the distribution $\chi(i_{\min})$ which assumes 1 at a state $i_{\min} := \arg\min_i\pi_i$
and zeros at all other states.
The normalized RMS deviation of a distribution $p$ in LJ$_N$ from $\pi$ is defined as
\begin{equation}
\label{rms}
d_{NRMS}(p,\pi) := \frac{\sqrt{\sum_i(p_i - \pi_i)^2} }{ \sqrt{\sum_i(\chi(i_{\min})_i - \pi_i)^2} }
\end{equation}

\begin{figure}[htbp]
(a)\includegraphics[width=0.8\textwidth]{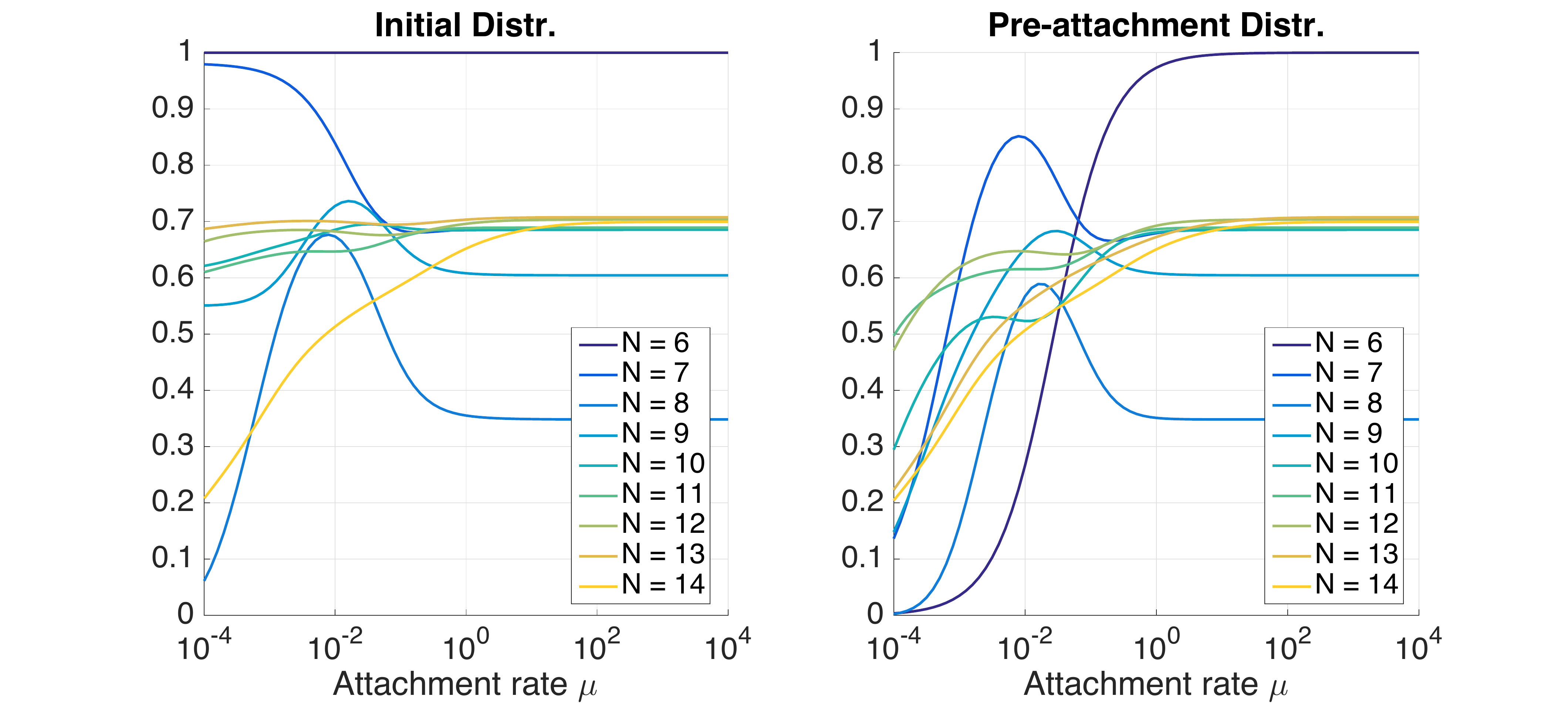}\\
(b)\includegraphics[width=0.8\textwidth]{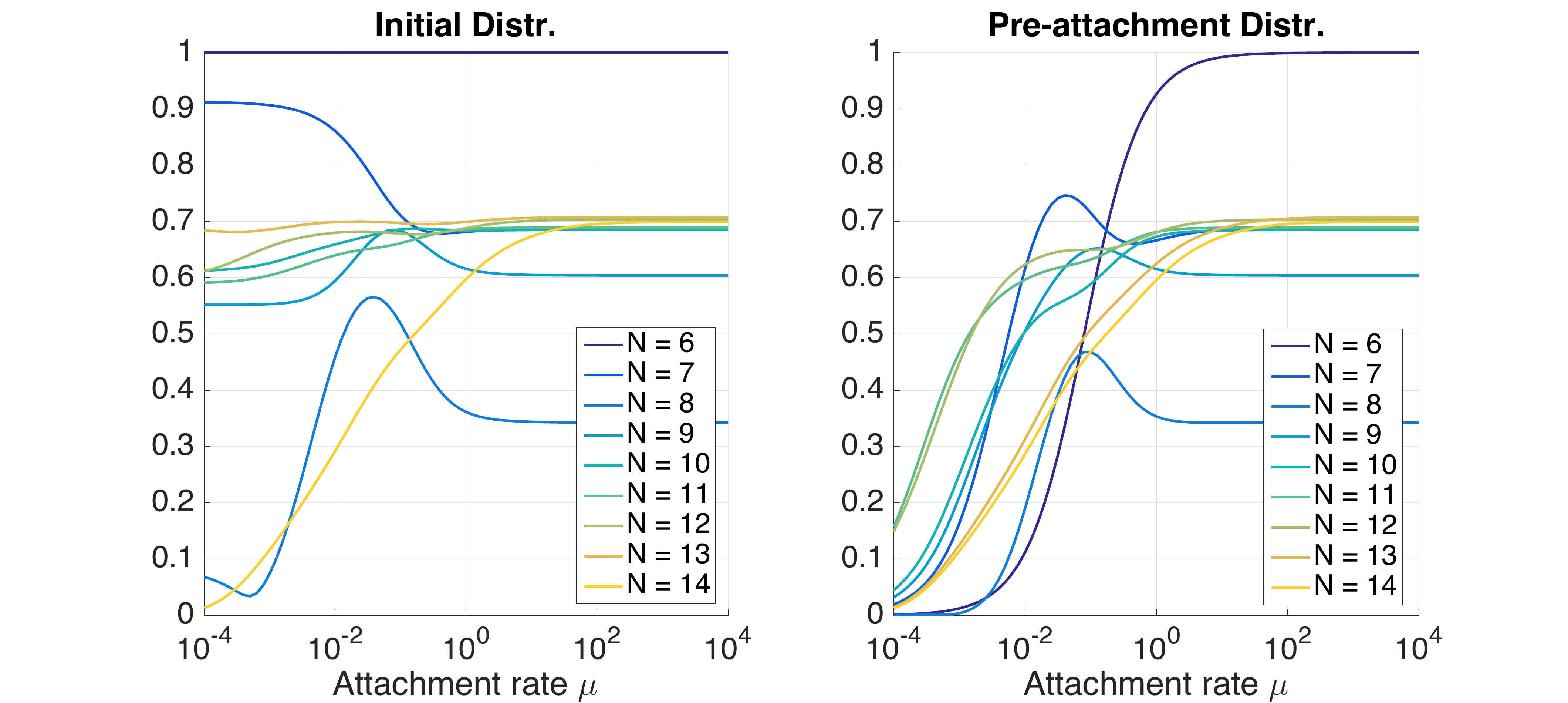}\\
(c)\includegraphics[width=0.8\textwidth]{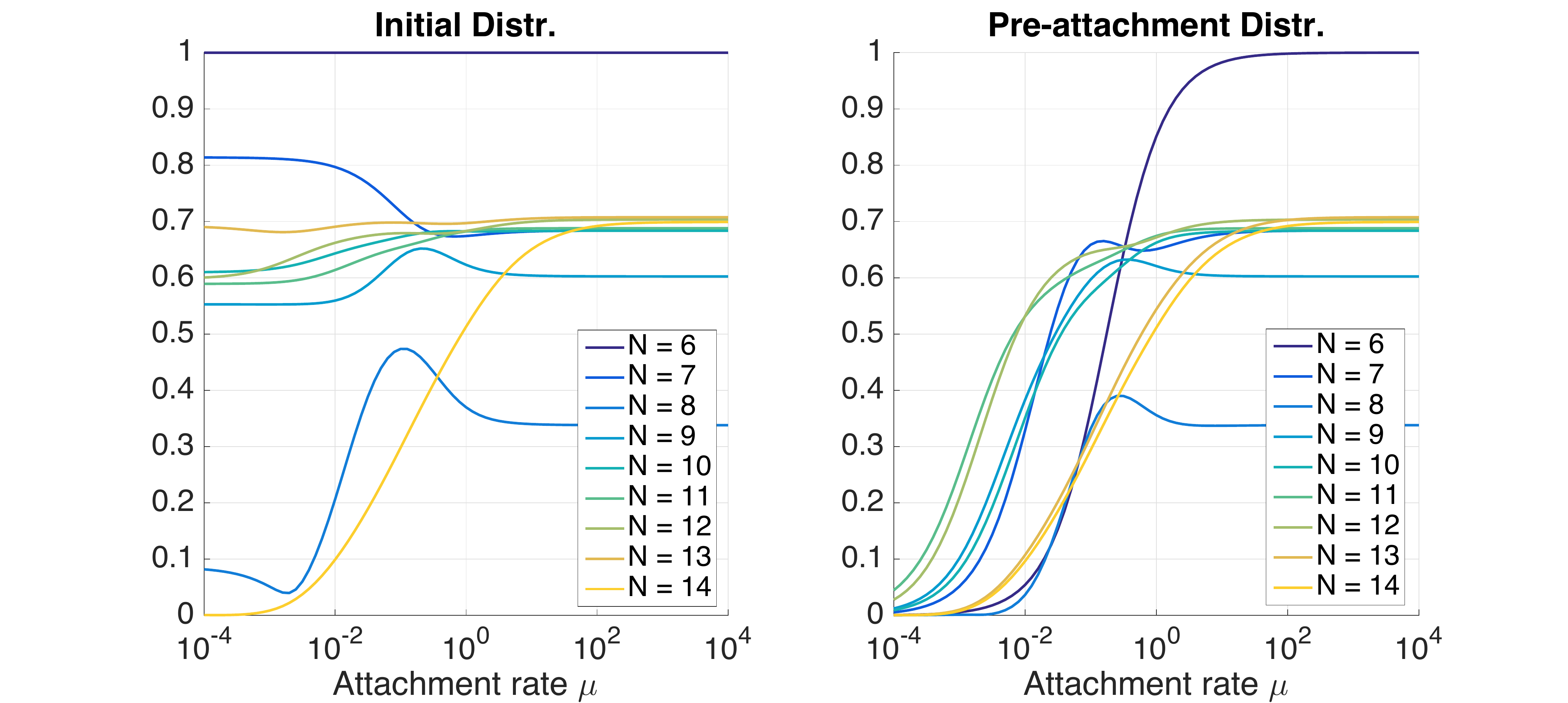}\\
\caption{The normalized RMS deviations of the expected initial and pre-attachment
distributions from the invariant distributions for $6\le N\le 14$.
(a): $\beta^{-1} =0.06$. (b): $\beta^{-1}=0.08$. (c): $\beta^{-1} = 0.10$.
} 
\label{fig:dev}
\end{figure}

The NRMS deviations of the expected initial and pre-attachment distributions from
the invariant distributions $\pi^N$ (Eq. \eqref{eq:pi}), $6\le N\le 14$, for $\beta^{-1}=0.06$, 0.08, and 0.10
are shown in Fig. \ref{fig:dev} (a),(b),(c) respectively. 
For all $N$, as one would expect, 
 $d_{NRMS}(p_0^N,\pi^N)$ and $d_{NRMS}(p_e^N,\pi^N)$
approach {$d_{NRMS}(b^N,\pi^N)$  (the normalized RMS deviation of the asymptotic distribution 
$b^N$ (Eq. \eqref{att}) from $\pi^N$) }as $\mu\rightarrow \infty$.
As $\mu\rightarrow 0$, $d_{NRMS}(p_0^N,\pi^N)$ and  $d_{NRMS}(p_e^N,\pi^N)$ approach 
$d_{NRMS}(\pi^{N-1}\Gamma^{N-1\rightarrow N},\pi^N)$ and zero respectively.
An interesting fact observed in Fig. \ref{fig:dev} is that the deviations 
$d_{NRMS}^N(p_0^N)$  for $N = 7$  and $9\le N\le 13$ are far from 0 for all attachment rates $\mu$. 
In particular, $d_{NRMS}(p_0^{13},\pi^{13})$
is nearly constant. 
This means that the attachment of a new atom throws invariant distributions for $N = 6$ and $8\le  N\le 12$
far away from the invariant distributions in $N = 7$ and $9\le N\le 13$ {respectively}, roughly as far as the distributions $b^N$. 
On the other hand, the global minima for $N = 8$ and $N=14$ are formed with probability one from the global minima 
of LJ$_7$ and LJ$_{13}$ respectively (Fig. \ref{fig:trans}).
This explains why the corresponding expected initial distributions are close to the invariant ones for low attachment rates $\mu$.
This effect is notably  stronger for LJ$_{14}$ because 
the global minimum of {LJ$_{14}$ is much deeper than all
other minima in LJ$_{14}$}, while the two deepest minima of LJ$_8$ have close values of energy.


\subsection{Structural analysis}
\label{sec:struct}
Fig. \ref{fig:redblue} and Table 2 suggest that the 6-atom octahedron M6(1) has a large icosahedral heritage that includes the global minima of
LJ$_N$, $12\le N\le 14$. In this Section, we make the concepts of icosahedral  or non-icosahedral packing more precise and 
quantify the structural transitions from icosahedral to non-icosahedral packings and vice versa during the attachment process.

\begin{figure}[htbp]
(a)\includegraphics[width=0.6\textwidth]{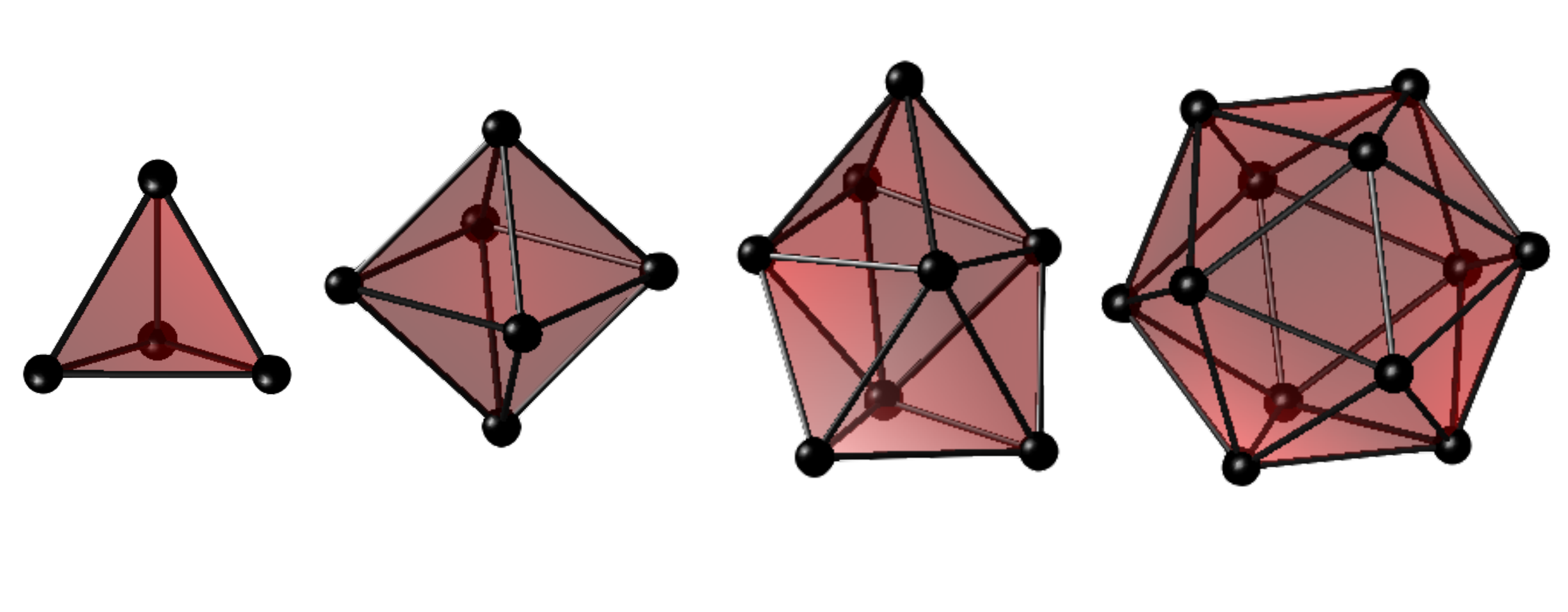}\\
(b)\includegraphics[width=0.6\textwidth]{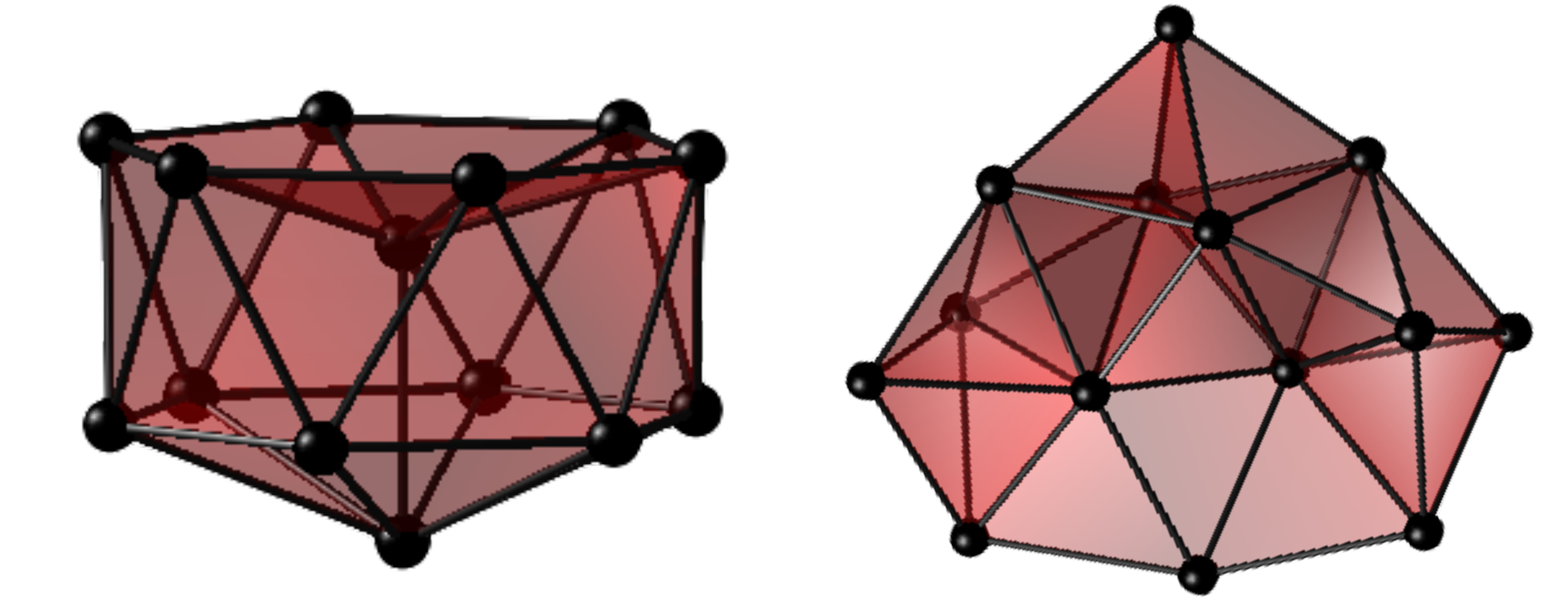}\\
\caption{(a): The main building blocks of LJ clusters (left to right): the regular tetrahedron, the octahedron, 
the M8(2) configuration with the point group $D_{2h}$ of order 8, and
the icosahedral shell.
(b): The two found minima of LJ$_{14}$ that do not consist of the building blocks above: (left to right) 
M14(43) can be split to two 
 hexagonal pyramids {rotated with respect to each other},
M14(3422) contains an atom (the one at the bottom) that is not a part of any tetrahedron.  
} 
\label{fig:struct}
\end{figure}

We will call a local minimum M$N(i)$ of LJ$_N$ \emph{icosahedral} if the following two conditions hold:
\begin{enumerate}
\item every atom in M$N(i)$ is a vertex of a tetrahedron, whose vertices are a subset of 4 atoms of LJ$_N$,  
and edges are of length $2^{1/6}(1+\delta)$, where $|\delta|\le\delta_1$;
\item  no two atoms in M$N(i)$ are at distances $2^{1/6}\sqrt{2}(1+\delta)$, where $|\delta|\le\delta_2$, or  $2^{1/6}d(1+\delta)$, where $|\delta|\le\delta_3$.
The number $d$ is the distance between the pairs of
atoms in M8(2) with 5 nearest neighbors, symmetric with respect to its symmetry plane (Fig. \ref{fig:struct} (a)), $d\approx1.269$. 
\end{enumerate}
Otherwise, we call a local minimum M$N(i)$ of LJ$_N$ \emph{non-icosahedral}. 
{We emphasize that our definition of icosahedral and non-icosahedral minima
refers to their packing rather than symmetry groups. Such a liberty is justified in the context of the study of aggregation, as any 
icosahedral minimum in the sense of our definition can be completed to a nearly 
regular icosahedron by the attachment of the right number of new atoms to the right places.}

This definition is easy to check {by a simple computer program}.  We set $\delta_1=\delta_3=0.1$ and $\delta_2=0.05$. 
The second condition renders minima such as M9(16), a tricapped octahedron (Fig. \ref{fig:nonico}), non-icosahedral.
In M9(16), every atom is a vertex of a tetrahedron; however, there is an 
octahedron in the middle.

Roughly speaking, the majority of local minima in Lennard-Jones clusters 
can be thought of being assembled out {of} building blocks shown in Fig. \ref{fig:struct} (a):
tetrahedron {(cap)}, octahedron, M8(2), and hollow icosahedral shell. These blocks can be distorted to avoid cavities/overlaps. 
For example, M9(5) and M9(9) are capped M8(2), M10(28) is a bicapped M8(2) (Fig. \ref{fig:nonico}), M13(1159) is a capped icosahedral shell.  
The numbers of icosahedral and some types of non-icosahedral minima are listed in Table 3. We did not split the types ``M6(1) (octahedron) and M8(2)"
as their ``signature" interatomic distances, $d\approx 1.269$ and $\sqrt{2}\approx1.414$, are close in comparison with our tolerances $\delta_i$, $i=1,2,3$.
The only two non-icosahedral minima that are not of any of {these} listed types are those of LJ$_{14}$ shown in Fig. \ref{fig:struct} (b): M14(43) consists of two 
{hexagonal pyramids} rotated by 30 degrees with respect to each other; M14(3422) has an atom that is not a part of any tetrahedron. 
While for $10\le N\le 14$ the numbers of icosahedral minima are less than those of non-icosahedral, 
one can check that the probabilities to find a cluster 
in an icosahedral minimum assuming the invariant distribution in LJ$_N$, $7\le N\le 14$, are very close to one.
\begin{table}[htp]
\caption{Structure of $N$-atom clusters. 
The column ``ico" contains the numbers of icosahedral local minima.
The column ``M6(1) or M8(2)" contains the numbers of local minima involving some interatomic
distances characteristic {of} the octahedron or the M8(2).
The column ``ico shell" contains the numbers of local minima involving the 12-atom icosahedral shell,
capped for LJ$_{13}$ and bicapped for LJ$_{14}$.
The column ``other" contains the numbers of non-icosahedral local minima of none of the above types.
}
\begin{center}
\begin{tabular}{|c|c|c|c|c|}
\hline
$N$ &  ico & M6(1) or M8(2) & ico shell & other\\
\hline
6 & 1 & 1 & 0 & 0\\
7 & 3 & 1 & 0 & 0\\
8 & 5 & 3 & 0 & 0\\
9 & 11 & 10 & 0 & 0\\
10 & 26 & 37 & 0 & 0\\
11 & 72 & 97 & 0 & 0\\
12 & 175 & {339} & 1 & 0\\
13 & 483 & 1026 & 1 & 0\\
14 & 1286 & 2842 & 5 & 2\\
\hline
\end{tabular}
\end{center}
\label{table3}
\end{table}%

Transition probabilities from icosahedral/non-icosahedral minima of LJ$_N$ to 
 icosahedral/non-icosahedral minima of LJ$_{N+1}$ as a result of attachment of a new atom are displayed in Table 4. 
 Evidently,  icosahedral minima tend to transition to icosahedral ones, 
 and non-icosahedral minima tend to transition to non-icosahedral ones. However, 
the transition probabilities from icosahedral to non-icosahedral minima and non-icosahedral to icosahedral ones are nonzero; 
the latter {probabilities} exceed the former by about an order of magnitude, and are significant for $11\le N\le 13$.

\begin{table}[htp]
\caption{Transition probabilities from icosahedral/non-icosahedral  local minima of LJ$_{N}$ to 
 icosahedral/non-icosahedral local minima of LJ$_{N+1}$. ``ico" and ``nico" abbreviate ``icosahedral" and ``non-icosahedral " respectively.  }
\begin{center}
\begin{tabular}{|c|c|c|c|c|}
\hline
$N\rightarrow N+1$ & $\mathbb{P}($ico $\rightarrow$ ico$)$ & $\mathbb{P}($ico $\rightarrow$ nico$)$ & 
$\mathbb{P}($nico $\rightarrow$ ico$)$ & $\mathbb{P}($nico $\rightarrow$ nico$)$\\
\hline
$6\rightarrow 7$ & 1 & 0 & 1.634e-4 & 9.998e-1 \\
$7\rightarrow 8$ & 9.993e-1 & 6.702e-4 & 2.315e-4 & 9.998e-1 \\
$8\rightarrow 9$ & 1 & 0 & 2.540e-4 & 9.997e-1 \\
$9\rightarrow 10$ & 9.997e-1 & 2.960e-4 & 1.118e-3 & 9.989e-1 \\
$10\rightarrow 11$ & 9.993e-1 & 7.112e-4 & 5.002e-3 & 9.950e-1 \\
$11\rightarrow 12$ & 9.935e-1 & 6.534e-3 & 7.440e-2 & 9.926e-1 \\
$12\rightarrow 13$ & 9.926e-1 & 7.436e-3 & {1.570e-1} &{ 8.430e-1} \\
$13 \rightarrow 14$ & 9.893e-1 & 1.071e-2 & 2.272e-1 & 7.728e-1\\
\hline
\end{tabular}
\end{center}
\label{table4}
\end{table}%

Tables 5 and 6 list the numbers of icosahedral and non-icosahedral minima in the distributions 
$a^N$ and $b^N$, $6\le N\le 14$, together with their probabilities.
The distributions $a^N$ contain both icosahedral and non-icosahedral clusters in comparable proportions for $12\le N\le 14$.
In contrast to this fact, the distributions $b^N$
{contain} primarily icosahedral minima.

Therefore, the  two kinds of processes, relaxation and attachment, involved in the aggregation process up to 14 atoms  lead to the formation of icosahedral clusters
for $N\ge 11$.
Relaxation does so because the global minima of LJ$_N$, $7\le N\le 14$, are icosahedral. 
Attachment favors icosahedral minima because icosahedral minima {transition primarily} to icosahedral ones, 
while non-icosahedral minima start to transition 
to both icosahedral and non-icosahedral ones with comparable probabilities  for $11\le N\le 13$. 

\begin{table}[htp]
\caption{The structure of local minima in the distributions $a^N$ (Eq. \eqref{att}). 
The columns ``ico" and ``nico" contain the numbers of icosahedral/non-icosahedral local minima respectively 
corresponding to nonzero entries in the distributions $b^N$, 
and the columns ``$\mathbb{P}$(ico)" and ``$\mathbb{P}$(nico)" contain their probabilities.
}
\begin{center}
\begin{tabular}{|c|c|c|c|c|}
\hline
$N$ & ico & $\mathbb{P}$(ico) & nico & $\mathbb{P}$(nico) \\
\hline
7 & 1 & 1.634e-4 & 1 & 9.998e-1\\
8 & 2 & 3.949e-4 & 3  & 9.996e-1 \\
9 & 9 & 8.221e-4 & 4  & 9.992e-1 \\
10 & 23 & 8.390e-4 & 22 & 9.992e-1 \\
11 & 68 & 4.336e-3 & 64 & 9.957e-1 \\
12 & 171 & 1.151e-1 & 233 & 8.492e-1 \\
13 & 475 & 3.251e-1 & 709 & 6.749e-1 \\
14 & 1264 & 4.828e-1 & 1987 & 5.172e-1\\
\hline
\end{tabular}
\end{center}
\label{table5}
\end{table}%

\begin{table}[htp]
\caption{The structure of local minima in the distributions $b^N$ (Eq. \eqref{att}). 
The columns ``ico" and ``nico" contain the numbers of icosahedral/non-icosahedral local minima respectively  
corresponding to nonzero entries in the distributions $b^N$, 
and the columns ``$\mathbb{P}$(ico)" and ``$\mathbb{P}$(nico)" contain their probabilities.
}
\begin{center}
\begin{tabular}{|c|c|c|c|c|}
\hline
$N$ & ico & $\mathbb{P}$(ico) & nico & $\mathbb{P}$(nico) \\
\hline
7 & 3 & 1 & 0 & 0 \\
8 & 5 & 9.991e-1 & 1 & 8.962e-4 \\
9 & 11 & 9.991e-1 & 2 & 8.955e-4 \\
10 & 25 & 9.990e-1 & 16 & 9.969e-4 \\
11 & 69 & 9.988e-1 & 58 & 1.165e-3 \\
12 & 171 & 9.982e-1 & 220 & 1.831e-3 \\
13 & 475 & 9.978e-1 & 687 & 2.203e-3 \\
14 & 1264 & 9.967e-1 & 1955 & 3.263e-3 \\
\hline
\end{tabular}
\end{center}
\label{table6}
\end{table}%

\section{Perspectives}
\label{sec:conclusion}
The aggregation/deformation LJ$_{6-14}$ network constructed and analyzed in this work is 
a model for an isothermal aggregation process, i.e., 
some amount of energy is taken away from to the cluster as it acquires a new atom in such a manner that the 
mean kinetic energy per atom remains constant. 
In this work, we had only one control parameter, the attachment rate $\mu$.
We assumed that the attachment time was an exponentially distributed random 
variable with a fixed parameter $\mu$ for all $N$. We did not allow detachments of atoms.
{Our analysis of this simple aggregation model showed that both processes taking place in the system, 
attachment and relaxation, promote icosahedral packing.}

{Our results encourage us to examine more sophisticated aggregation models, in particular, enabling detachments,
in our future work.  
Figs. \ref{fig:distr1} -- \ref{fig:distr3}  suggest the conjecture }
that the primary mechanism of the formation of the 13-atom icosahedron is from the global minimum of LJ$_{14}${,}
the capped icosahedron:
the ``cap" atom detaches from the icosahedron. 
Each atom on the surface of the 13-atom icosahedron
has 6 nearest neighbors, which makes it extremely stable.  
This would explain the notable peaks in the mass spectra in \cite{exp1,exp4}
at $N=13$. Presumably, a similar mechanism takes place for other clusters with magic numbers of atoms. 

Besides allowing detachments, 
the study of aggregation processes by means of stochastic networks 
can be continued in several {other} directions. First, one can continue building LJ$_{6-N}$ aggregation/deformation networks
for $N>14$. Due to the exponential growth of the number of local minima in LJ$_N$ with $N$ (Eq. \eqref{eq:LJmin}),
it will be necessary to use some kind of importance sampling on the set of local minima, 
e.g.,  the basin hopping method \cite{wales110,wales0}.
For example, Wales's datasets for LJ$_{38}$ \cite{web} and LJ$_{75}$\footnotemark[1] contain $100\thinspace000$ and $593\thinspace320$ 
local minima respectively, while the  {predicted } numbers of local minima in them according to Eq. \eqref{eq:LJmin}
are of the orders of $10^{14}$ and $10^{31}$ respectively. 
\footnotetext[1]{Courtesy of David Wales.}

Second, one can consider a non-isothermal  aggregation and make the attachment rate  $\mu$
dependent on the current number of atoms in the cluster. { For example, one can imagine 
a fixed number of interacting macroscopic particles (e.g., ball-shaped macromolecules) 
that are allowed to self-assemble in a small closed container filled with solvent (e.g., see experiments conducted with microgel balls in \cite{short2}).}

Finally, { our methodology of the study of aggregation process of Lennard-Jones particles
by means of stochastic networks is transferable to the study of self-assembly of particles 
interacting according to other kinds of potentials.
The dream of design by self-assembly inspired research 
on the self-assembly of micron-size particles interacting according to a short-range potential 
\cite{short1,short2,holmes,holmes2016}, limited to a fixed number of
particles so far. Allowing new particles to arrive  at a controlled rate  and regulating the temperature will 
upgrade the ability to obtain desired configurations of particles. }

The present work can be considered as the first step toward the goal
of generating desired types of clusters by means of controlled aggregation/self-assembly.

\section*{Acknowledgement}
This work was partially supported by the NSF grant DMS1554907 and the NSF REU grant DMS1359307 at the University of Maryland, College Park. 


\begin{thebibliography}{00}

\bibitem{short1}
N. Arkus, V. Manoharan and M.~P. Brenner, 
Minimal Energy Clusters of  Hard Spheres with Short Ranged Attractions, 
Phys Rev Lett, {\bf 103},118303 (2009).


{
\bibitem{varsize1}
F. Baletto, A. Rapallo, G. Rossi, and R. Ferrando,
Dynamical effects in the formation of magic cluster structures,
Phys. Rev. B {\bf 69}, 235421 (2004)
}

\bibitem{becker_karplus}
O.~M. Becker, and M. Karplus, 
The topology of multidimensional potential energy surfaces: Theory and application to peptide structure and kinetics,
J. Chem. Phys. {\bf 106} (4), 1495 (1997)


{
\bibitem{varsize2}
F. Calvo, D. Schebarchov, and D. J. Wales, 
Grand and Semigrand Canonical Basin-Hopping,
J. Chem. Theory Comput. {\bf 12}, 902 - 909 (2016)
}

{
\bibitem{mydata}
\begin{verbatim}https://www.math.umd.edu/~mariakc/lennard-jones.html\end{verbatim}
}

\bibitem{cve}
M. Cameron and E. Vanden-Eijnden, 
Flows in Complex Networks: Theory, Algorithms, and Application to Lennard-Jones Cluster Rearrangement, 
J. Stat. Phys., {\bf 156}, 3, 427-454 (2014)

\bibitem{C-eigen}
M.~K. Cameron, Metastability, Spectrum, and Eigencurrents of the Lennard-Jones-38 Network, 
J. Chem. Phys., {\bf 141}, 184113 (2014)

\bibitem{CG} 
M.~K. Cameron and T. Gan, {Spectral analysis and clustering of large stochastic networks. Application to the Lennard-Jones-75 cluster.} 
Molecular Simulation, {\bf 42}, 16,  1410-1428 (2016)

\bibitem{beta3}
J.~M. Carr and D.~J. Wales, 
Folding Pathways and Rates for the Three-Stranded  -Sheet Peptide Beta3s using Discrete
Path Sampling,
J. Phys. Chem. B, {\bf 112}, 8760-8769 (2008)

\bibitem{wales-thermo}
J.~P.~K. Doye, D.~J. Wales, and M.~A. Miller,
Thermodynamics and the global optimization of Lennard-Jones clusters,
J. Chem. Phys. {\bf 109}, 8143 (1998)

\bibitem{wales38}
J.~P.~K. Doye, D.~J. Wales, and M.~A. Miller,
The double-funnel energy landscape of the 38-atom Lennard-Jones cluster,
J. Chem. Phys. {\bf 110}, 6896 (1999)

\bibitem{string1}
 W. E, W. Ren, and E.~Vanden-Eijnden, String method for the study of rare events, 
 Phys. Rev. B: {\bf 66}, 052301 (2002)
 
 \bibitem{string2}
 W. E, W. Ren, and E. Vanden-Eijnden, 
 Simplified and improved string method for computing the minimum energy paths in barrier-crossing events,
 J. Chem. Phys.: {\bf 126},164103 (2007)

\bibitem{xiang1}
W. E and X. Zhou, 
The gentlest ascend dynamics, 
Nonlinearity, {\bf 24}, 1831-1842 (2011)

\bibitem{exp1}
O. Echt, K. Sattler, and E. Recknagel, 
Magic Numbers for Sphere Packings: Experimental Verification in Free Xenon Clusters, Phys. Rev. Lett., {\bf 47}, 16, 1121-1124 (1981) 


\bibitem{exp2}
O. Echt and O. Kandler and T. Leisner and W. Mlechle and E. Recknagel, 
Magic Numbers in Mass Spectra of Large van der Waals Clusters
J. Chem. Soc. Faraday Trans., {\bf 86}, 2411 (1990)

\bibitem{exp3}
 J. Farges and M. F. de Feraudy and B. Raoult and G. Torchet, 
 Structure and temperature of rare gas clusters in a supersonic expansion,
 Surface Science {\bf 106}, 95, (1981)


\bibitem{helix}
S.~N. Fejer and D.~J. Wales, 
Helix Self-Assembly from Anisotropic Molecules,
Phys. Rev. Letters, {\bf 99}, 086106 (2007)

\bibitem{xiang2}
W. Gao, J. Leng, and  X. Zhou, 
Iterative minimization algorithm for efficient calculations of transition states,
J. Comp. Phys., {\bf 309},  69Ð87 (2016)

\bibitem{exp4} 
I. A. Harris and R. S. Kidwell and J. A. Northby, 
Structure of charged argon clusters formed in free jet expansion,
Phys. Rev. Lett., {\bf 53}, 2390 (1984)

\bibitem{exp5} 
I. A. Harris and K. A. Norman and R. V. Mulkern and J. A. Northby, 
Icosahedral structure of large charged argon clusters,
Chem. Phys. Lett., {\bf 130}, 316  (1986)

\bibitem{dimer}
G. Henkelman and H. Jonsson, 
A dimer method for finding saddle points on high dimensional potential surfaces using only first derivatives, 
J. Chem. Phys., {\bf 111}, 7010 (1999). 

\bibitem{holmes}
M. Holmes-Cerfon, S.~J. Gortler, M.~P. Brenner, A geometrical approach to computing free-energy landscapes from short-ranged potentials,
Proc. Natl. Acad. Sci. {\bf 110} , 1,  E5 - E14 (2013)

\bibitem{holmes2016}
M. Holmes-Cerfon, 
Sticky-sphere clusters, 
Annual Reviews of Condensed Matter Physics, In press (expected {\bf 8}, March 10, 2017).

\bibitem{neb}
H. Jonsson, G. Mills, K. W. Jacobsen, 
Nudged Elastic Band Method for Finding Minimum Energy Paths of Transitions, 
in Classical and Quantum Dynamics in Condensed Phase Simulations, 
Ed. B. J. Berne, G. Ciccotti and D. F. Coker, 
385 (World Scientific, 1998).

\bibitem{kakar}
S. Kakar, O. Bjoerneholm, J. Weigelt, A. R. B. de Castro, L. Troeger, R. Frahm, and T. Moeller, 
Size-dependent K-edge EXAFS study of the structure of free Ar clusters,
Phys. Rev. Lett. {\bf 78}, 9, 1675-1678 (1997)
%

\bibitem{kovalenko}
S. I. Kovalenko and D. D. Solnyshkin and E. T. Verkhovtseva and V. V. Eremenko, 
Experimental detection of stacking faults in rare gas clusters
Chem.
Phys. Lett., {\bf 250}, 309  (1996)

\bibitem{langer}
J.~S. Langer, 
Statistical Theory of the Decay of Metastable States, 
Ann. Phys. {\bf 54}, 258-275 (1969)

\bibitem{mf}
V.~A. Mandelshtam and P.~A. Frantsuzov, 
Multiple structural transformations in Lennard-Jones clusters: Generic versus size-specific behavior,
J. Chem. Phys. {\bf 124}, 204511 (2006)

{
\bibitem{short2}
G. Meng, N. Arkus, M.P. Brenner and V. Manoharan, The Free Energy Landscape of Hard Sphere Clusters, Science {\bf  327},  560 (2010)
}

\bibitem{LJ7}
M.~A. Miller and D.~J. Wales,
Isomerization dynamics and ergodicity in Ar 7,
J. Chem. Phys. {\bf 107}, 8568 (1997)

{
\bibitem{munro}
L.~J. Munro and D.~J. Wales, 
Defect migration in crystalline silicon,
Phys. Rev. B, {\bf 59}, 3969-3980 (1999)
}

\bibitem{nocedal}
J. Nocedal and S.~J. Wright, Numerical Optimization, Second Edition, Springer, USA, 2006 

\bibitem{picciani}
M. Picciani, M. Athenes, J. Kurchan, and J. Taileur, 
Simulating structural transitions by direct transition
current sampling: the example of LJ38, 
J. Chem. Phys. {\bf 135}, 034108 (2011)

{
\bibitem{stillinger}
F.~H. Stillinger, Exponential multiplicity of inherent structures,
Phys. Rev. E, {\bf 59}, 1, 48-51 (1999)

\bibitem{wallace}
D.~C. Wallace, Statistical mechanics of monatomic liquids, 
Phys. Rev. E, {\bf 56}, 4, 4179-4186 (1997)
}

\bibitem{wales110}
D.~J. Wales and J.~P.~K. Doye,
Global Optimization by Basin-Hopping and the Lowest Energy Structures of Lennard-Jones Clusters Containing up to 110 Atoms,
J. Phys. Chem. A, {\bf 101}, 5111-5116, (1997)

\bibitem{wales0}
 D.~J. Wales, {\it Discrete Path Sampling}, 
Mol. Phys., {\bf 100} (2002), 3285-3306

\bibitem{wall}
B.~W. van de Waal, No Evidence for Size-Dependent Icosahedral $\rightarrow$ fcc Structural Transition in Rare-Gas Clusters,
Phys. Rev. Lett., {\bf 76}, 7, 1083-1086  (1996)

\bibitem{wales-book}
D.~J. Wales, ``Energy Landscapes: Applications to Clusters, Biomolecules and Glasses", Cambridge University Press, 2003

\bibitem{wales-landscapes}
 D.~J. Wales, {\it Energy landscapes: calculating pathways and rates}, 
International Review in Chemical Physics, {\bf 25}, 1-2,  237-282 (2006)

\bibitem{wales-short}
D.~J. Wales, Energy Landscapes of Clusters Bound by Short-Ranged Potentials,
J. Chem. Phys. and Phys. Chem., {\bf 11}, 12, 2491-2494 (2010)

\bibitem{web}
\begin{verbatim} http://www-wales.ch.cam.ac.uk/CCD.html \end{verbatim}

\bibitem{du-zhang}
J. Zhang and Q. Du, Shrinking dimer dynamics and its applications to saddle point search,
SIAM J. Numer. Anal., {\bf 50}, 4, 1899-1921 (2012)

\end{thebibliography}
\end{document}